\documentclass[%
reprint,
superscriptaddress,
showpacs,
nofootinbib, nobibnotes, 
 amsmath,
 amssymb,
 aps,
 prl
]{revtex4-2}

\usepackage{microtype}
\usepackage{blindtext}
\usepackage{dcolumn}
\usepackage{bm}
\usepackage{color, soul, colortbl} 
\usepackage[colorlinks=true,
						linkcolor=blue,
						urlcolor=blue,
						citecolor=blue,
						bookmarks=true,
						pdfborder={0 0 0}]{hyperref}

\usepackage{xcolor}
\usepackage{mdframed} 

\newcommand{\SI}[1]{\textcolor{blue}{{\it SI Appendix} #1}} 

\usepackage{graphicx,epsfig}
\usepackage{subfigure}
\usepackage{amsmath,amsfonts}
\usepackage{xfrac}
\usepackage{enumerate}
\usepackage[overload]{empheq}
\usepackage{mathtools}
\usepackage{cases} 
\usepackage{cancel} 

\usepackage{booktabs} 

\begin{document}
\title{Social interactions affect discovery processes}

\author{Gabriele Di Bona}
\email[To whom correspondence should be addressed. E-mail: ]{g.dibona@qmul.ac.uk}
\affiliation{School of Mathematical Science, Queen Mary University of London, London E1 4NS, United Kingdom}
\author{Enrico Ubaldi}
\affiliation{SONY Computer Science Laboratories, Paris, 6, rue Amyot, 75005, Paris, France}
\affiliation{MindEarth, 2502, Biel/Bienne, Switzerland}
\author{Iacopo Iacopini}
\affiliation{Department of Network and Data Science, Central European University, 1100 Vienna, Austria}
\author{Bernardo Monechi}
\affiliation{SONY Computer Science Laboratories, Paris, 6, rue Amyot, 75005, Paris, France}
\author{Vito Latora}
\affiliation{School of Mathematical Science, Queen Mary University of London, London E1 4NS, United Kingdom}
\affiliation{Dipartimento di Fisica ed Astronomia, Università di Catania and INFN, I-95123 Catania, Italy}
\affiliation{Complexity Science Hub, Josefst \"{a}adter Strasse 39, A 1080 Vienna, Austria}
\author{Vittorio Loreto}
\affiliation{SONY Computer Science Laboratories, Paris, 6, rue Amyot, 75005, Paris, France}
\affiliation{Sapienza Univ. of Rome, Physics Dept., Piazzale Aldo Moro 2, 00185 Rome, Italy}
\affiliation{Complexity Science Hub, Josefst \"{a}adter Strasse 39, A 1080 Vienna, Austria}

\date{\today}

\begin{abstract} 
Our network of acquaintances determines how we get exposed to ideas, products, or cultural artworks (books, music, movies, etc.). Though this principle is part of our common sense, little is known about the specific pathways through which our peers influence our discovery processes and our experience of the new. Here, we fill this gap by investigating a data set containing the whole listening histories of a large, socially connected sample of users from the online music platform \emph{Last.fm}. We demonstrate that users exhibit highly heterogeneous discovery rates of new songs and artists and that their social neighborhood significantly influences their behavior. More explorative users tend to interact with peers more prone to explore new content. We capture this phenomenology in a modeling scheme where users are represented by random walkers exploring a graph of songs or artists and interacting with each other through their social links. Even starting from a uniform population of agents (no natural differences among the individuals), our model predicts the emergence of strong heterogeneous exploration patterns, with users clustered according to their musical tastes and propensity to explore. We contend our approach can pave the way to a quantitative approach to collective discovery processes.
\end{abstract}
\maketitle

{\bf In this work, we investigate the collective exploration of conceptual spaces and the ensuing experience of novelties. We focus, in particular, on the space of musical artworks, and we show that the network of contacts affects the propensity of individuals to explore new content, as observed empirically. We find that users tend to cluster in communities according to their musical tastes and propensity to explore. We present a general theoretical framework that correctly reproduces key statistical features of the empirical data, giving insights into the social mechanisms that affect how people collectively explore the space of possibilities. We contend that our results can help shape the innovation paths for individuals and groups.}
\newline

In our everyday life, we are continuously exposed to novel ideas, new information, innovative cultural and technological products, and so on~\cite{ziman2003innovation, arthur2009technology, mcnerney2011role, hofstra2020diversity}. 
Understanding the subtle balance between the exploration of new opportunities and the exploitation of what we already know is fundamental to unveil how we build our knowledge and set of skills~\cite{erdi2013prediction, kim2017dynamic, tacchella2020language, zhou2020growth}.
Such a task is even more challenging if we consider that we live in a more and more interconnected society~\cite{albert2002statistical, newman2003structure, latora_nicosia_russo_2017} and we do not explore the world alone. Indeed, we are constantly influenced by our peers, directly or indirectly~\cite{hodas2014simple,centola2020behavior,ternovski2020social,iacopini2021dual}.

The recent increase in quantity and quality of digital traces 
has unlocked the possibility of tracking individual exploration trajectories in systems and processes as diverse as online music consumption~\cite{zangerle2014nowplaying,brost2019music,schedl2021recommendation}, food purchases~\cite{di2018sequences, aiello2019large, schulz2019structured}, Twitter post creation~\cite{weng2015topicality}, and code development~\cite{monechi2017waves}. 
These opportunities have allowed, among other things, to characterize the typical exploration patterns of individual users and to investigate the drivers that lead to a novelty, defined as the first time a user adopts or consumes a given content~\cite{wu2007novelty, tria2014dynamics, rzhetsky2015choosing, monechi2017waves}.
Although social interactions play a crucial role in all such systems~\cite{iacopini2020discovery}, a thorough data-driven understanding of their impact on individual and collective exploration trajectories is still lacking. In this article, we try to fill this gap by studying how social connections and communities affect the exploration patterns of different users of online (social) music platforms. 
Our work is based on a unique data set that, differently from those used in previous studies~\cite{lastfm360k1k,cantador2011lastfm,dror2012yahoo,mcfee2012million,hauger2013million,vigliensoni2017music,schedl2016lfm}, contains information on both the whole listening histories and the social connections of a large and connected sample of users from the online music listening platform \emph{Last.fm}.
This platform is particularly suitable for our purposes, since it specifically encourages interactions among users and pushes them to explore new songs based on the rich metadata attached to each track~\cite{lastfm_about}. 
As such, the data set we use represents the ideal testbed to:  \emph{i}) characterize the exploration and discovery dynamics of each user, \emph{ii}) measure the impact of social interactions, and \emph{iii}) provide a structured view of the conceptual (or musical, in this case) space being explored by the agents~\cite{maccallum2012evolution,siew2019cognitive, lynn2020humans, lydon2021hunters}.
%
%

A first analysis of the data set reveals that user exploration behaviors are very heterogeneous. Moreover, we find that users with high discovery rates, the so-called explorers, tend to be connected with similar users in the social network, indicating the presence of homophily~\cite{mcpherson2001birds}. 
To get a better insight into the interplay between the topology of the social network and the different propensities of the users to explore, we introduce a multiagent model in which the agents simultaneously explore a space of contents. Our model is based on the so-called {\em urn model with semantic  triggering}~(UMST)~\cite{tria2014dynamics}, which is already capable of reproducing some statistical features of the empirical exploration trajectories, e.g., the Zipf's, Heaps' and Taylor's  laws~\cite{loreto2016dynamics,monechi2017waves,iacopini2018network,tria_2020_taylor}. 
UMSTs have been recently adapted both to model the evolution of a social network~\cite{ubaldi2021emergence} and to investigate peer effects in discovery dynamics~\cite{iacopini2020discovery}. While the former model approach does not deal with content exploration~\cite{ubaldi2021emergence}, the latter lacks semantics in the space being explored~\cite{iacopini2020discovery}. Here, we explicitly consider a space endowed with a complex network of semantic relations between artists as given by similarity. Agents independently navigate this network with a reinforcement mechanism while interacting with each other through an underlying social network, progressively enlarging their space of possibilities.
%
%
The collective nature of the dynamics allows us to correctly capture and reproduce the key empirical findings at a local and global level. At the local level, one observes an assortative arrangement of explorers, while at the global level, communities of people sharing similar music tastes emerge. The ensemble of these results provides important insights into the empirical interplay between the individual and the collective experience of the new.

\section*{Results}
\subsection*{The social network of \emph{Last.fm}}




\emph{Last.fm} is an online digital music streaming platform born in 2002, famous for logging all listening activities, known as \emph{scrobbles}, of its users.
We crawled the \emph{Last.fm} platform using its API, collecting all the listening sequences and social connections of a group of 4836 users found, growing a breadth-first search sub-graph from a random seed. We end up with 335'375'125 unique streaming events complete of metadata and timestamps, with a total of 6'972'047 unique tracks authored by 958'732 artists (more information in \emph{Materials and Methods}).

\begin{figure*}[ht!]
    \centering
    \includegraphics[width=\textwidth]{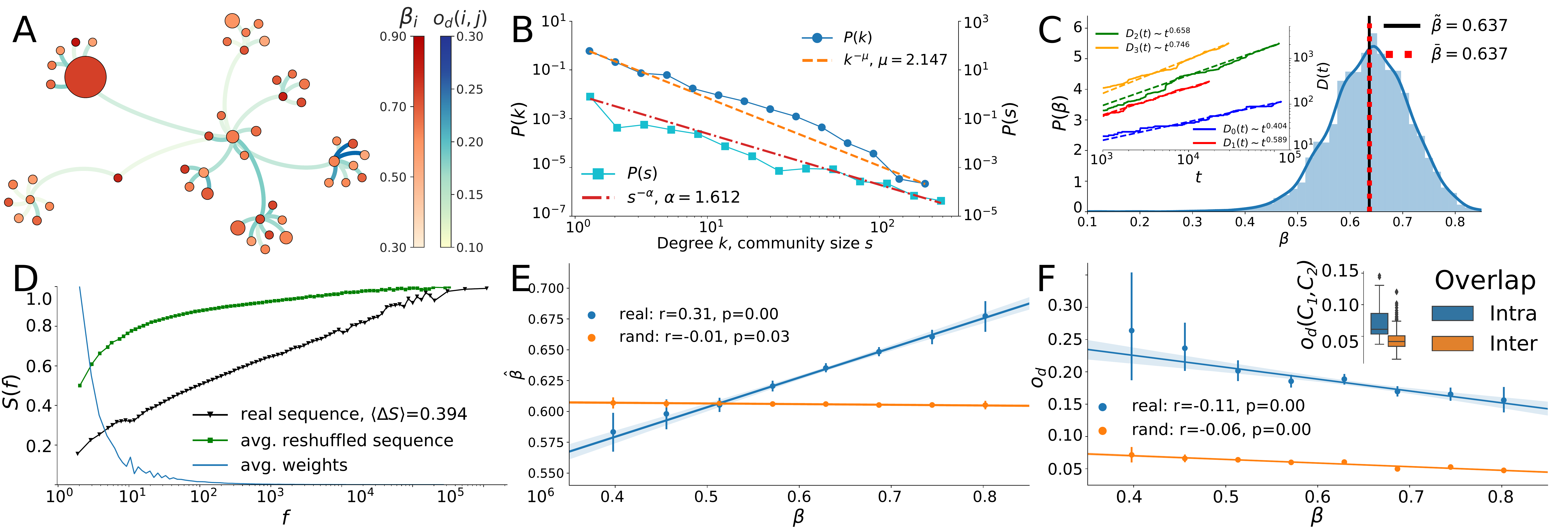}
\caption{
    \label{fig:data set}
    \textbf{ Statistical analysis of the \emph{Last.fm} data set.} 
    \textbf{(A)} Sub-graph of the social network $\mathcal{G}_S$ of the analyzed sample. The node size is proportional to its betweenness centrality, while color intensity is proportional to its Heaps' exponent $\beta$ (the redder, the higher). The color intensity of a link $e_{ij}$ is proportional to the dynamical overlap $o_d(i,j)$ of the two nodes (the bluer, the higher).
    \textbf{(B)} Degree distribution $P(k)\sim k^{-\mu}$ (blue dots) and the community size distribution $P(s) \sim s^{-\alpha}$ (cyan squares) of $\mathcal{G}_S$, both fitted as a power-law with exponents $\mu \approx 2.147$ (orange dashed line) and $\alpha \approx 1.612$ (red dash-dot line).
    \textbf{(C)} Heaps' exponent distribution $P(\beta)$ for all the users in the data set. We also show the average of the distribution $\bar \beta \approx 0.64$ (black line), and the global Heaps' exponent $\tilde \beta \approx 0.64$ measured on the global sequence $\mathcal{S}$ (red dotted line). In the inset we show the Heaps' laws and fitted exponents of four random users, plotting the number $D(t)$ of distinct artists in the user's sequence $\mathcal{S}$ as a function of its length $t$. 
    \textbf{(D)} Normalized Shannon entropy $S(f)$ of the sequences of listened artists as a function of the artist frequency $f$ (black line), compared to the entropy $\tilde S(f)$ measured on the reshuffled sequences (green line). We also show the weights for each frequency $f$ (cyan line) used to compute $\langle\Delta S\rangle \approx 0.394$ (more details in \emph{Materials and Methods}).
    \textbf{(E)} Assortativity of the Heaps' exponent: the average exponent $\hat{\beta}$ measured over the neighbors of a node with exponent $\beta$ is plotted as a function of $\beta$. Results from the original network $\mathcal{G}_S$ (in blue) are compared to those of a randomized social network using a configuration model (in orange).
    \textbf{(F)} Correlation between the dynamical overlap $o_d$ of each user with its neighbors against the Heaps' exponent $\beta$ of the user. Both the original (blue) and the randomized social network (orange) are shown. In the inset, we compare the intra- (blue) and inter-community (orange) dynamical overlap distribution $o_d(C_1,C_2)$, where in the former $C_1=C_2$, and in the latter $C_1\neq C_2$.
  }
\end{figure*}

In Fig.~\ref{fig:data set}(A), we show a local snapshot of the users' social network $\mathcal{G}_S$, where nodes are colored according to their exploration propensity (i.e., their Heaps' exponent $\beta_i$ defined below, the redder, the higher), their size is proportional to their betweenness centrality, and the link color intensity is proportional to the dynamical overlap (a measure of similarity between users based on their listening sequences, see below for the definition). 
As shown in Fig.~\ref{fig:data set}(B), $\mathcal{G}_S$ features a scale-free degree distribution $P(k)\sim k^{-\mu}$, where $k$ is the degree of a user, i.e., the number of users in the sample followed on \emph{Last.fm}, and $\mu \approx 2.15$, with average degree $\langle k \rangle \approx 7.88$. The network is highly reciprocal (i.e., both the $i\to j$ and the $j\to i$ links exist), since 98.5\% of links are reciprocal. $\mathcal{G}_S$ can be considered a typical small-world network~\cite{watts1998collective}, featuring, in fact, a small-world coefficient $\sigma\approx1.35$~\cite{humphries2008smallworldness}, with a relatively low characteristic path length (5.68) and high clustering coefficient (0.15). The sample is representative of the main statistics of the whole network, since we find a Pearson correlation coefficient of value $r\approx0.839$ ($p < 0.0001$) between the nodes' degree $k$ in the sample and the number of real friends they have on the platform (see \SI{1} for details).
The network also features a community structure~\cite{fortunato2010community}---i.e., nodes are arranged in tightly connected groups that are weakly linked one to the other---that we detected using the Louvain algorithm~\cite{blondel2008fast}. In Fig.~\ref{fig:data set}(B), we show the community size distribution $P(s)$, finding 31 communities with more than ten users, with an average size of 150 users.

%
%
\subsection*{Heaps' laws}
To quantify the exploration rate of new artists, we measure the Heaps' law exponent from the sequence $\mathcal{S}_i$ of artists listened by each user $i$ in the sample, i.e., each node $i$ of the social network $\mathcal{G}_S$. The Heaps' law links the number $D(t)$ of distinct elements that are found in a sequence of $t$ elements to a power-law behavior, namely $D(t) \propto t^\beta$, where $0 < \beta \le 1$ is the \emph{Heaps' exponent}~\cite{heaps1978information}. This ubiquitous law has been found to hold in a variety of data sets logging the evolution of different systems~\cite{tria2014dynamics, monechi2017waves}. 
Moreover, under certain hypotheses, the Heaps' law is related to the Zipf's law~\cite{serrano2009modeling,lu2010zipf,gerlach2013stochastic}, stating that the frequency $f$--rank $R$ distribution of the elements in a sequence $\mathcal{S}$ decays as $f\propto R^{-1/\beta}$ for large ranks $R$
~\cite{Zipf,tria2014dynamics} (see \SI{3} for details).

In our analysis, the Heaps' exponent represents a natural proxy to measure the discovery rate of each user, which we hypothesize directly related to the propensity to search for new content during the discovery process. We estimate it as $\beta_i = \log D(t_i) / \log t_i$, i.e., the average slope in the log-log plane of the Heaps' law of user $i$ (see \SI{2} for further details). In Fig.~\ref{fig:data set}(C), we observe how the Heaps' exponents $\beta$ are heterogeneously distributed among the population. Some users are in fact more open to consume new music (higher $\beta$), while some others are more inclined to exploit already known tunes (lower $\beta$). The inset of Fig.~\ref{fig:data set}(C) shows how the Heaps' exponent is estimated from the listening sequences of four different users. We also note that a large fraction of the population features a discovery rate around the average $\bar\beta\approx0.64$. Interestingly, this is the same value of the Heaps' exponent $\tilde \beta \approx 0.64$ found by measuring the Heaps' exponent on the whole global sequence $\mathcal{S}$, i.e., the single sequence obtained by merging, in temporal order, the records of all the users.
%
%
\subsection*{Semantic correlations}
The semantic correlations among the occurrences of the artists within the listening histories of users represent another interesting observable. To estimate these correlations, we compute, for each user $i$ and each artist $a$ listened by $i$, a modified Shannon entropy $S_i^{a}(f)$ introduced in ~\cite{tria2014dynamics}, where $f$ denotes the number of streams of artist $a$ in the user's sequence $\mathcal{S}_i$ (see \emph{Material and Methods} for additional details).
This measure quantifies to which extent the occurrences of an artist $a$ in the listening sequence of a user $i$ are clustered, or, in other words, whether listening to a given artist increases the probability of listening to it again shortly after.
To test the statistical significance of this measure, we compare it with the randomized counterpart $\tilde S_i^a(f)$, i.e., the Shannon entropy of each artist $a$ listened by the user $i$ computed after reshuffling the sequence $\mathcal{S}_i$. 
In Fig.~\ref{fig:data set}(D) we plot the entropy $S(f)$ averaged over all the artists and over all users with frequency $f$ in $\mathcal{S}_i$. The empirical data show a lower entropy (higher clustering) of the same-artist streaming events when compared to the reshuffled case. To quantify the difference between the two distributions, we weight each frequency $f$ with the global popularity $w_f$, namely the number of times we find an artist with frequency $f$ in the different user sequences. The resulting weighted difference is $\langle\Delta S\rangle = \sum_f w_f (\tilde{\mathcal{S}}(f) - \mathcal{S}(f)) / \sum_f w_f \approx 0.394$. 
Overall, the presence of significant semantic correlations implies that users tend to listen to music semantically close to the recent plays. Still, they also experience new content according to the Heaps' law from time to time.

\subsection*{The influence of the social network}
We now shift the attention from the individual to the collective level. To this end, we focus on the relation between the position of users in the social network and their respective exploration strategies. As we qualitatively show in Fig.~\ref{fig:data set}(A), users tend to interact with people featuring a similar discovery rate and musical tastes.
We quantitatively explore this assortativity in Fig.~\ref{fig:data set}(E), where we observe a positive correlation between the Heaps' exponent $\beta_i$ of a user $i$ and the average exponent $\hat{\beta}_i=\langle \beta_j\rangle_{j \sim i}$ of its neighbors, featuring a Pearson correlation coefficient of $r\approx0.31$ ($p<0.0001$). To test its significance, we measure the same correlation on a network obtained by randomly rewiring the edges of $\mathcal{G}_S$, obtaining in this case $r\approx-0.01\,(p<0.05)$.
%
This evidence is a clear sign of homophily based on the discovery rate $\beta_i$: explorers (exploiters), i.e., people with higher (lower) $\beta$ exponent tend to form clusters with other explorers (exploiters).

The influence of the social network can also be measured by looking at the dynamical overlap $o_d(i,j)$ of a pair of users $i$ and $j$, i.e., the fraction of common artists they listen to. This is calculated as $o_d(i,j) = \tilde{\mathbf{v}}_i \cdot \tilde{\mathbf{v}}_j$, where $\cdot$ is the scalar product and $\tilde{\mathbf{v}}_i$ is the vector of the normalized frequency distribution of artists listened by user $i$.
If we average the dynamical overlap of a node $i$ with its neighbors, that is $o_d(i) = \langle o_d(i,j) \rangle_{j\sim i}$, we can then compare it against its discovery rate $\beta_i$, as shown in Fig.~\ref{fig:data set}(F). We find that the average dynamical overlap $o_d(i)$ of a user negatively correlates with its discovery rate $\beta_i$ (Pearson's $r\approx-0.11$ against $r\approx-0.06$ in the rewired network). 
This result reveals that explorers tend to interact slightly more with people sharing different musical tastes, thus enlarging the set of artists and genres they are exposed to. On the contrary, exploiters preferably surround themselves with people sharing similar tastes, limiting their chances to explore new content. The evidence that, in both cases, the average dynamical overlap is much higher than the one calculated on the rewired network represents a general signature of homophily (friends share similar tastes).

Moving away from the local scale, we then test if the self-organization of users based on their tastes does also hold at the community level. For each pair of communities $C_1,\,C_2$ in the social network, we compute their inter-community dynamical overlap as the average overlap of all possible pairs inside $C_1,\, C_2$, that is $o_d(C_1,\,C_2) = \langle o_d(i,j)\rangle_{i \in C_1, j \in C_2, i \neq j}$. When $C_1=C_2=C$, we obtain the intra-community overlap $o_d(C) = \langle o_d(i,j)\rangle_{i,j\in C,i \neq j}$. 
In the inset of Fig.~\ref{fig:data set}(F) we compare the intra- and inter-community overlap distributions of communities with at least ten users. We show that the average intra-community overlap (0.074) is significantly larger than the average inter-community one (0.039). In other words, users tend to stay in communities of people sharing the same musical tastes.\newline


So far, we have found three main results concerning the collective exploration of the spaces of artists: 
\emph{i)} users feature different propensity to explore new content, showing a heterogeneous distribution of discovery rate $\beta$; 
\emph{ii)} users tend to interact and cluster with others sharing a similar propensity to explore new content (positive assortativity of exponent $\beta$); 
\emph{iii)} social connections are mainly established between groups of people sharing similar tastes (higher dynamical overlap within the communities).
In the following, we will try to reproduce these results leveraging a multi-agent model of content spaces exploration. Before this, the last ingredient we need is a tool to measure, from the empirical data, the semantic structure of the content space to be explored, as we show in the next paragraph.

\subsection*{Topology of the content space}
\begin{figure*}
    \centering
    \includegraphics[width=0.95\linewidth]{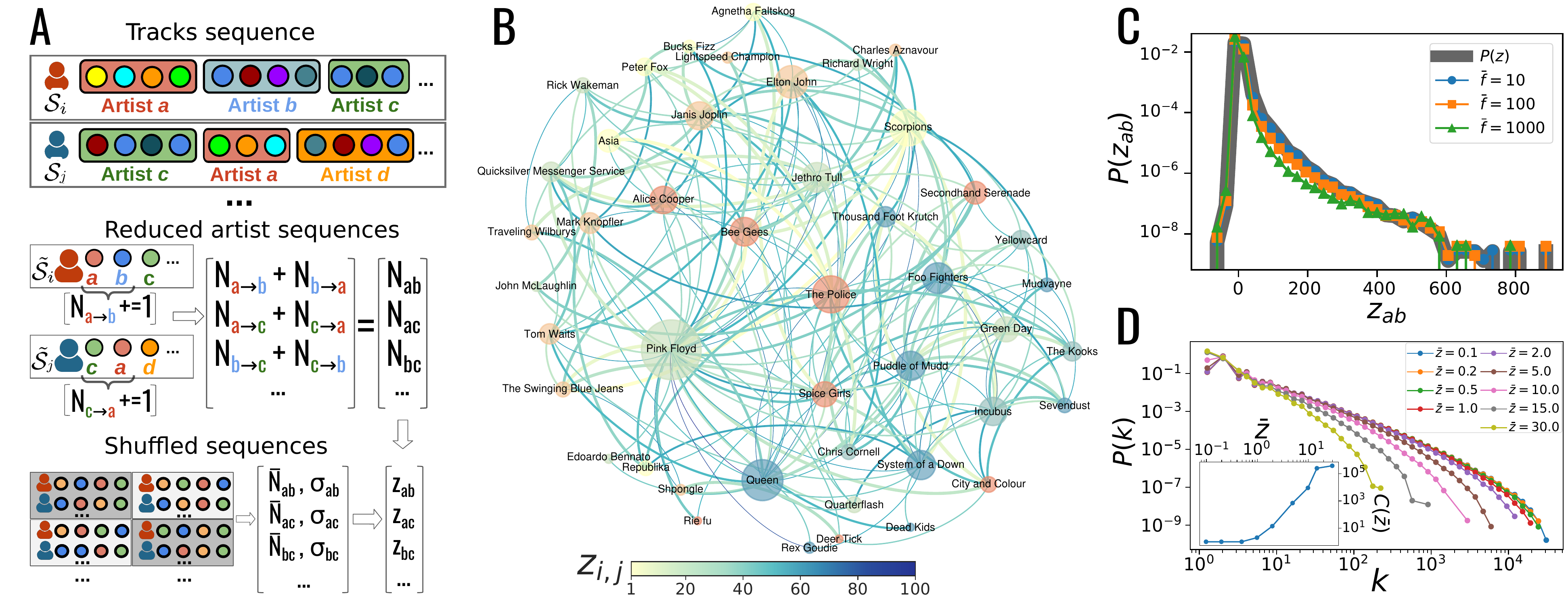}
\caption{
    \label{fig:artists_network}
    \textbf{ From streams of songs to the creation of the content space.} 
    %
    \textbf{(A)} Illustration of the method. First, for each user $i$'s sequence $\mathcal{S}_i$ we compress all consecutive appearances of an artist, say  $a$, into a single occurrence of $a$, forming, in this way, the reduced sequence $\tilde{\mathcal{S}}_i$. Then, we count the number $N_{ab} = N_{a \to b}+N_{b \to a}$ of pairs of consecutive artists $a$ and $b$ in both orders in $\tilde{\mathcal{S}}_i$. We repeat the procedure $Q=100$ times after reshuffling $\tilde{\mathcal{S}}_i$, calculating $N^q_{ab}$ for each reshuffle $q$. We thus evaluate the z-score $z_{ab} = (N_{ab}-\bar{N}_{ab})/\sigma_{ab}$, where $\bar{N}_{ab}$ and $\sigma_{ab}$ are respectively the average and standard deviation of all $N^q_{ab}$, for each pair of artists $a$ and $b$. Artists with overall frequency less than $\bar{f}$ are disregarded, while we draw a link between two artists $a$ and $b$ if $z_{ab} > \bar{z}$.
    \textbf{(B)} Snowball-sampled snapshot of the neighborhood of \emph{Pink Floyd} in the space of artists. Node sizes are proportional to their degree, while their color depends on the community of belonging. The color of the edges denotes their weight according to the z-scores (see \emph{Materials and Methods}), the bluer, the larger.
    \textbf{(C)} Comparison of the original z-score distribution (gray) with those obtained using different thresholds on the artist frequencies $\bar{f}$.
    \textbf{(D)} Degree distribution of the artists' network for different values of the threshold $\bar{z}$. The inset shows the number of connected components of the network as a function of the threshold $\bar{z}$.
  }
\end{figure*}

The temporally-ordered sequences, containing the streams of songs for each user, allow us to characterize their overall discovery processes and extract a semantic structure of the content space. Because of the scarcity of meaningful musical features attached to the records in the data set, we quantify the similarity between artists from sequences of listening events.  In doing this, we are hypothesizing the existence of an underlying semantic structure of the musical space that users navigate to consume content~\cite{iacopini2018network}.
To this end, we construct a bootstrap-like statistics for the number of times, $N_{ab}$, a user listens to artist $a$ and $b$ sequentially (irrespectively of the order of $a$ and $b$). We do that by computing the expectation $\overline N_{ab} = \langle N^q_{ab}\rangle_q$ and the standard deviation $\sigma_{ab} = \rm{std}(N^q_{ab})$ of the $N_{ab}$ occurrences over $q=1,\ldots,100$ shuffles of the users sequences, as shown in Fig.~\ref{fig:artists_network}(A). Note that we filter out from this analysis all artists with a total number of appearances $f < \bar f$ to remove noise. We can then compute how much each $a \leftrightarrow b$ transition is over-expressed using the z-score $z_{ab} = (N_{ab} - \overline N_{ab}) / \sigma_{ab}$. We thus obtain a proximity network between the artists, in which a link is drawn if two artists $a$ and $b$ feature $z_{ab} > \bar z$, where $\bar z$ is a fixed threshold (see \emph{Materials and Methods} for details).
A snapshot of such space of contents is shown in Fig.~\ref{fig:artists_network}(B), where a snowball sample of the neighborhood of the artist \emph{Pink Floyd} is shown. As one can see, artists belonging to the same music genre fall within the same community. In Fig.~\ref{fig:artists_network}(C-D), we also show that both the z-score distribution and the degree distribution of the network encoding the space of content are stable with respect to the choice of different values of $\bar z$ and $\bar f$.

The procedure defined above transforms the universe of items in the data set---the artist IDs---into a weighted network $\mathcal{G}_C$ where we can measure distances between nodes. This representation also allows us to look at the sequence of records differently, i.e., as the exploration of a proximity network of artists. In particular, knowing the semantic distance between artists, we can measure the propensity of users to explore items falling outside their comfort zones and their willingness to accept recommendations (from others) not strictly meeting their current musical tastes.
This mechanism will be a crucial ingredient of the model introduced in the next section.


\subsection*{Modeling collective exploration}
\begin{figure}[tb]
  \centering
  \includegraphics[width=\linewidth]{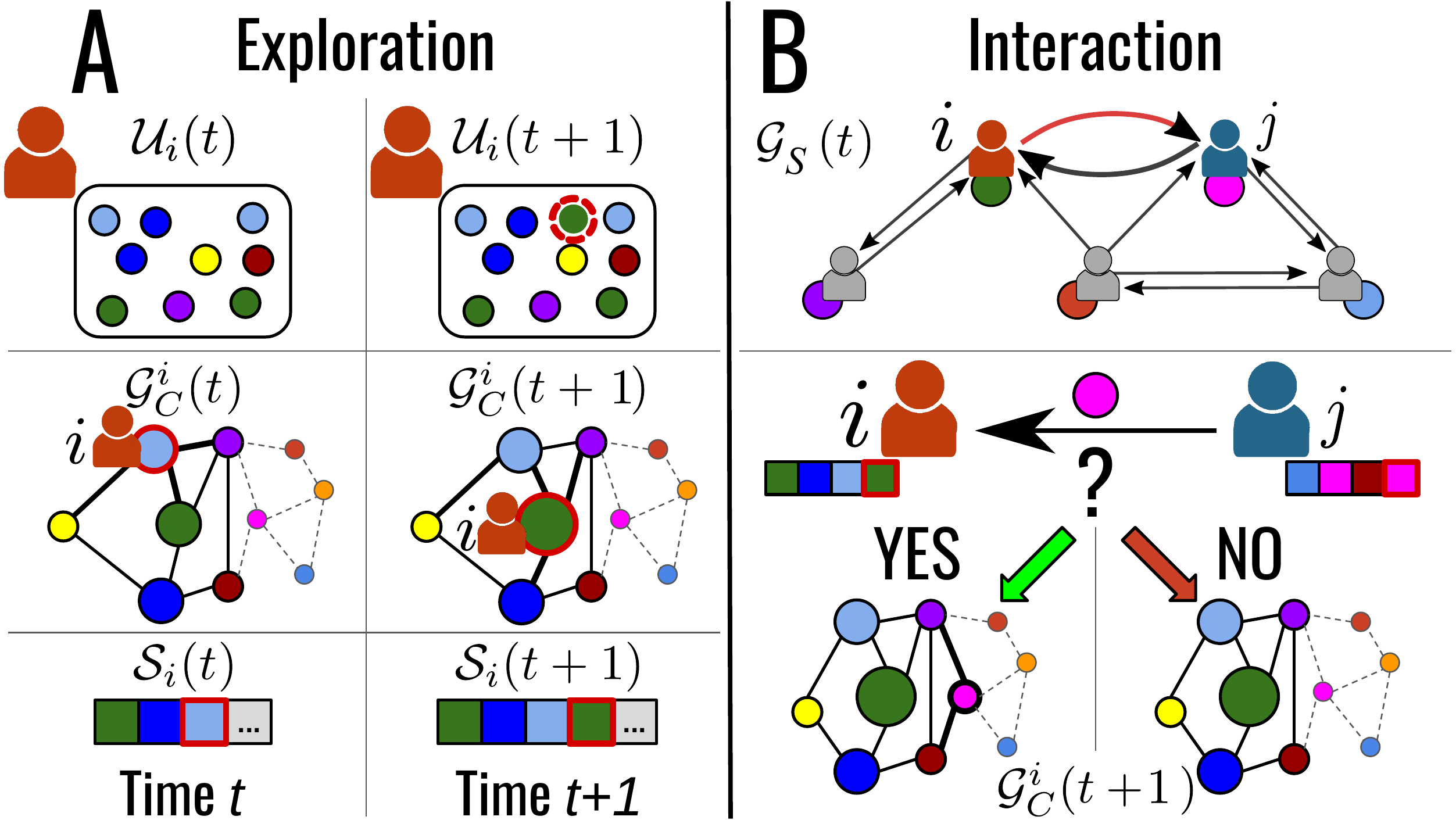}
  \caption{
    \label{fig:model}
    \textbf{Model definition.} 
    \textbf{(A)} \emph{Exploration step.} The agent $i$ extracts a colored ball from its urn $\mathcal{U}_i(t)$ (top left), each color representing a different item. Each ball in the urn has a probability of being extracted, at time $t+1$, depending on whether its color is semantically correlated to the recent history of the agent $i$. In the example, cyan is the last color extracted and added to the sequence $\mathcal{S}_i(t)$ at time $t$ (bottom left). Two colors are semantically correlated if they are neighbors in the conceptual network $\mathcal{G}_C^i$ of agent $i$ (middle left). With this rule, the colors yellow, green, and purple are semantically correlated to cyan, and they are marked with bold links in the network. So, for the extraction at time $t+1$, all balls of these three colors will have assigned weight $1$, while all other balls will have a weight $\eta \le 1$. Node sizes in the network represent the multiplicity of the corresponding color in the urn $\mathcal{U}_i(t)$. After the new color is chosen, in this case green, it is stored in $\mathcal{S}_i(t+1)$ (bottom right), $i$ moves to the corresponding node in $\mathcal{G}_C^i$ (middle right), and $\rho = 1$ new green balls are added to the urn (top right). 
    %
    %
   \textbf{(B)} \emph{Interaction step.} In the top panel, we represent the directed network of contacts, $\mathcal{G}_S(t)$, where $i$ and $j$ are neighbors. Green and pink nodes represent the last chosen colors for $i$ and $j$, respectively. After the exploration step, the agent $i$ randomly selects one of its neighbors to interact with, i.e., in this specific case, agent $j$ (top). The agent $i$ may accept the last visited item (pink) of $j$ with a probability depending on the semantic distance of the item from its experience and on its propensity to explore new content. In case $i$ accepts the pink item, it adds it in its urn $\mathcal{U}_i(t+1)$  and changes the structure of $\mathcal{G}_C^i$ (bottom left). If the item is too far from $i$'s tastes or $i$ is unwilling to accept contents outside its experience, the interaction does not lead to any change in $\mathcal{G}_C^i$ (bottom right).
  }
\end{figure}
To better understand the interplay between individual exploration and the effects brought by the social network, we now present an agent-based model. We build upon the individual exploration-exploitation dynamics introduced in the \emph{Urn Model with Semantic Triggering} (UMST)~\cite{tria2014dynamics}. In the UMST, the dynamics of an individual, or, from now onwards, an agent, is modeled as random extractions of colored balls from an urn $\mathcal{U}$ to form a sequence of events $\mathcal{S}$. The urn $\mathcal{U}$ represents the \emph{space of possibilities}, i.e., the set of possible choices the agent can make in the future. The space of possibilities includes the so-called \emph{actual space}, i.e., the subset of colors already extracted by the agent and stored in the sequence $\mathcal{S}$. In addition to the actual space, the urn contains the so-called \emph{adjacent possible space}, which consists of all those colors that are one step away from the actual space~\cite{packard1988adaptation,langton1990computation,kauffman1996investigations}. In the original UMST, the concept of proximity between colors is modeled through the definition of semantic relations between groups~\cite{tria2014dynamics}.
The idea is that the agent can realize only a subset of all the possibilities at any given time, preferentially extracting balls semantically related to the most recent ones.
In particular, at every extraction, balls in $\mathcal{U}$ whose color is semantically related to the last drawn color keep their unitary weight. In contrast, all the other balls get---temporarily---a weight $\eta \le 1$ (with $\eta=1$ we recover the classic \emph{Urn Model with Triggering}---UMT---also found in Ref.~\cite{tria2014dynamics}). Then, the agent draws a random ball with a probability proportional to these weights. The selected ball is put back in the urn with $\rho$ additional copies of it (reinforcement step). Finally, if this ball has never appeared in $\mathcal{S}$, $\nu+1$ brand-new balls are added to the urn (novelty-triggering step). 

Here, we extend the UMST model to account for peers' influence and the exploration of a shared conceptual space. Concretely, we allow $N$ agents to both explore a shared network of items $\mathcal{G}_C$, and interact with each other by exchanging information via a social network of contacts $\mathcal{G}_S$. All agents are bound to the same evolutionary rules with the same parameters, as described in the following.
%
An urn governs the basic exploration dynamics of each agent, where, differently from the classic UMST, semantic relations are introduced via $\mathcal{G}_C$. This space is, in general, a directed and weighted network, where link weights represent the strength of the semantic relation between pairs of items (depending on the context, these could be ideas, molecules, genomes, technological products, etc.). Interactions happen through the social network $\mathcal{G}_S$, itself a directed and weighted graph, whose links $i \to j$, of weight $w_{ij}$ that may generally change over time, represent the propensity of a node $i$ to follow and copy the tastes of $j$. In this context, the adjacent possible for an agent $i$ is the subset of items (nodes) in $\mathcal{G}_C$ that are neighbors of the elements in $i$'s actual space, which is enriched by both $i$'s personal exploration history and interactions with other users.
At the beginning of the simulation, each agent $i$ sits on a node $x_i(t=0)$ of the conceptual space (see \SI{4} for details about the initialization procedures). We assign to each agent an urn $\mathcal{U}_i(t=0) = \{x_i(t=0)\}$, representing the initial portion of the conceptual space that the agent can explore. The social network is then initialized, with weights $w_{ij}=1$ for each link $i\to j$, zero otherwise. 
In our simulations, we consider the same $N=4836$ agents as in the crawled \emph{Last.fm} data set, using their social relationships for $\mathcal{G}_S$ and the previously discussed proximity network between artists for $\mathcal{G}_C$.
Moreover, we work in the system intrinsic time $t$, i.e., each listening event increases $t\to t+1$.
As summarized in Fig.~\ref{fig:model}, each time step is composed of two processes. An agent $i$, active at time $t$, \emph{i}) independently explores its space of possibilities by randomly sampling a token from its urn $\mathcal{U}_i(t)$, and, \emph{ii}) interacts with one of its neighbors to query for recommendations. 
The choice of the active agent $i$ at time $t$ can either be random or, as in the following, the same as found in temporally-ordered sequences of the empirical data set.

\emph{Exploration step.} 
During the time step $t$ in which the agent $i$ is active, a token $x_i(t)$ is randomly drawn from $\mathcal{U}_i(t)$ with probability depending on the semantic relations with $i$'s recent history. It is important to remark that, at each time step, $\mathcal{U}_i(t)$ only contains balls representing items of the subset $\mathcal{G}_C^i(t)$ of the shared conceptual network $\mathcal{G}_C$, that is $i$'s space of possibilities. This subset comprises all items already seen by the agent $i$, i.e., its actual space, and a subset of neighbors in $\mathcal{G}_C$ of the items in $i$'s actual space, as well as the accepted items recommended by other agents. The recent history is represented by the last item visited by $i$ and the one possibly received from another agent during $i$'s last interaction step (see below). The weights used in the extraction from $\mathcal{U}_i(t)$ are unitary for all balls of items semantically related to the recent history of $i$, and $\eta \le 1$ for all others items (see \emph{Materials and Methods}). Notice that $i$ can choose the same element drawn in its last draw.
After the extraction, $i$ puts the ball back in its urn together with $\rho$ copies of it (reinforcement step). Additionally, if it is the first time that $i$ sees the chosen item $x_i(t)$---i.e., $x_i(t) \notin \mathcal{S}_i(t)$---then $i$ expands its adjacent possible space by adding $\nu+1$ new tokens semantically close to $x_i(t)$ into its urn with multiplicity one (novelty-triggering step). 
For more details on the sampling procedure, see \emph{Materials and Methods}.
%
%
%
%
These two mechanisms account for the repeated listening of known artists by a user (exploitation) and for the expansion of the space of possibilities every time a user experiences something new (exploration). Fixing $\eta$, the ratio between the values of the two other parameters $\rho$ and $\nu$ sets the relative weight of exploration and exploitation. 
As we will show later, social interactions represent a key ingredient to reproduce the assortativity of the exploration rates and the clusterization of individuals with similar tastes.
%
%
%

\emph{Interaction step.} 
After the exploration step, agent $i$ randomly selects another agent $j$ among its neighbors in $\mathcal{G}_S$ with a probability distribution given by the weights $w_{ij}$. To mimic the dynamics of an online listening platform like \emph{Last.fm}, $i$ has the possibility to see the last token $x = x_j(t)$ listened by $j$. We estimate $i$'s potential interest in $x$ with the semantic distance of $x$ from its $c_i$-core, i.e., the set of the first $c$ tokens ranked by frequency found in the sequence $\mathcal{S}_i(t)$. Agent $i$ actually adds $x$ into its urn with probability $P(x,\mathcal{S}_i(t),c_i,\varepsilon)$, where $\varepsilon$ is a noise factor that mimics $i$'s imperfect capability of estimating the distance of items from its $c_i$-core (see \emph{Materials and Methods} for details).
Finally, the model also allows the social network to co-evolve while interactions between agents occur. In this version, the weight of the link $i \to j$ is increased (decreased) by $+\Delta$ ($-\Delta$) if $i$ accepts (rejects) the suggestion.
In the following, for the sake of simplicity, we focus on the case in which $\mathcal{G}_S$ does not change over time ($\Delta = 0$). We refer to \SI{6} for the results on an evolving $\mathcal{G}_S$, starting from a fully connected network.
\subsection*{Testing the model}
\begin{figure*}
    \centering
    \includegraphics[width=0.975\textwidth]{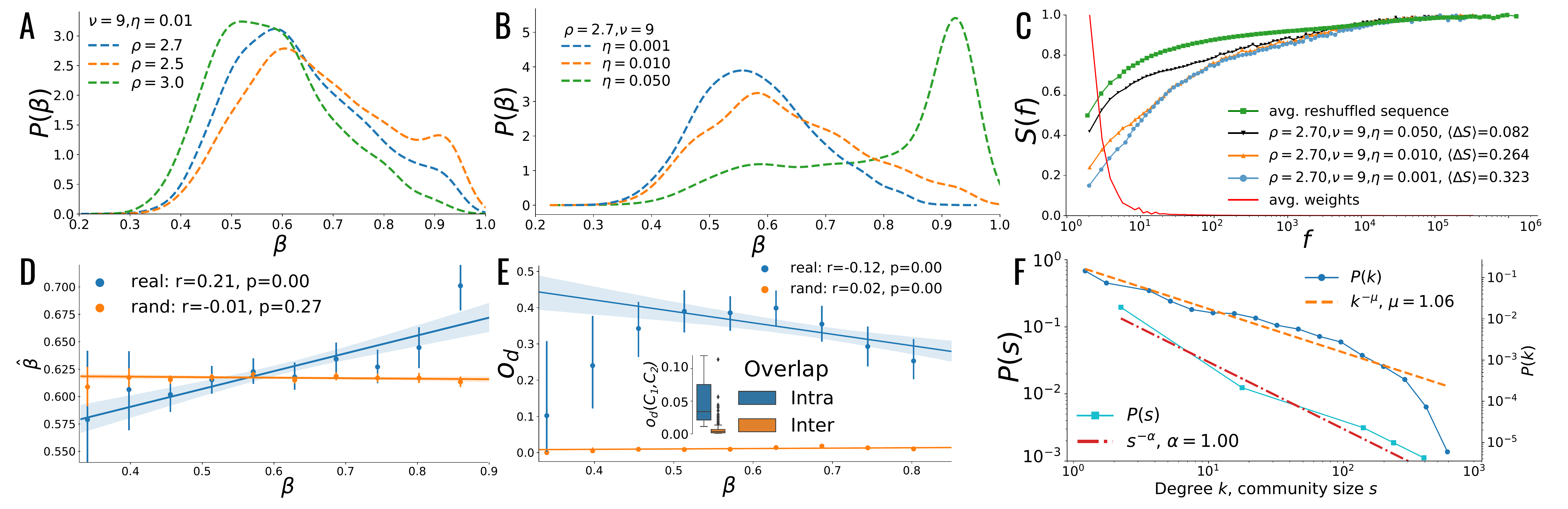}  

\caption{
    \label{fig:simulations}
    Analysis of the key observables monitored during the simulations.
    \textbf{(A)} Heaps' exponents distribution $P(\beta)$ in the reference simulation ($\rho=2.7,\, \nu=9,\, \eta=0.01$) (blue dashed line), compared to two cases with a lower ($\rho=2.5$, orange line) and higher ($\rho=3.0$, green line) value of $\rho$. 
    \textbf{(B)} Heaps' exponents distribution $P(\beta)$ in the reference simulation (orange dashed line), compared to two cases with a lower ($\eta=0.001$, blue line) and higher ($\eta=0.05$, green line) value of $\eta$.
    \textbf{(C)} Shannon entropy distribution of the occurrences of all artists in the user sequences in the same simulations showed in (B), compared to the Shannon entropy calculated on the reshuffled sequences of the reference case (green line). We also show the average weights used for each frequency $f$ (red line) used to compute the different $\langle\Delta S\rangle$ (refer to \emph{Materials and Methods} for details).
    \textbf{(D)} Scatter plot between the user Heaps' exponents $\beta$ and the average neighbors' Heaps' exponent $\hat \beta$, as found in the reference simulation (blue line) and when randomly rewiring the social network links using a configuration model (orange line). We observed the presence of assortativity between users through their exploration rates.
    \textbf{(E)} The average dynamical overlap $o_d$ of the neighbors' as a function of the node's $\beta$ Heaps' exponent, as found in the reference simulation (blue line) and when randomly rewiring the social network links using a configuration model (orange line). In the inset, we show the average dynamical overlap distribution between users in the same community (blue box) and between different communities (orange box).
    \textbf{(F)} The $P(k)\propto k^{-\mu}$ degree distribution (blue circles) and the $P(s)\propto s^{-\alpha}$ community-size distribution (cyan squares) as found when letting the social network co-evolve during the model simulation using the reference set of parameters (see the \emph{Materials and Methods} for details). We also show the power-law fitting of the two giving $\mu\approx1.06$ (orange line) and $\alpha\approx1.00$ (red line).
  }
\end{figure*}
%
%
This section tests to which extent our model reproduces the empirical observables. To this end, we analyze the outcome of the simulated exploration process with different values of the parameters. For computational reasons, we analyze simulation runs for about 33.5 million evolution steps instead of the total 335 million present in the data set (see \SI{7} for a comparison with the latter case). This way, we can \emph{i}) inspect the influence of each parameter on the model outcomes, and, \emph{ii}) select a reference set of parameters that reproduces the empirical data. 
We select such configuration by looking at the simulations that have similar semantic correlations to the empirical data and that minimize the Kullback-Leibler $KL$ divergence between the empirical and synthetic Heaps' exponents distributions $P(\beta)$ in the population (see the \emph{Materials and Methods} for details). This way, we are not putting any constraints on the other empirical observables, such as assortativity or taste overlaps. In our case, the reference combination of parameters is given by $\rho=2.7, \nu=9, \eta=0.01$.
%
%
%
%
%

First, we check in Fig.~\ref{fig:simulations}(A,B) how the distribution $P(\beta)$ of Heaps' exponents changes, varying respectively only $\rho$ and $\eta$. We find that both higher values of $\rho$ and lower values of $\eta$ decrease the average Heaps' exponent. This result means that the higher the reinforcement strength $\rho$, the higher the probability of extracting already known tokens. Similarly, with lower values of $\eta$, the agent has higher chances to move only within semantically close tokens, thus usually ignoring all other possibilities even with large values of the triggering parameter $\nu$.
These plots also show that the model can reproduce heterogeneous $P(\beta)$ distributions with different mean, even though all the agents share the same set of parameters and evolution rules. This evidence represents our first finding: collective exploration allows agents in our model to have heterogeneous exploration propensities.

Next, in Fig.~\ref{fig:simulations}(C), we show that with low values of $\eta$, the synthetic individual sequences of tokens feature strong semantic correlations as in the empirical case. In particular, the lower the parameter $\eta$, the lower the Shannon entropy $S(f)$ is, which implies the presence of stronger semantic correlations in the sequences. Hence, the evolution rules that promote the extraction of items semantically correlated to the last extracted one generate highly correlated sequences of tokens, featuring low values of the Shannon entropy $S(f)$.
%
%

Moreover, in Fig.~\ref{fig:simulations}(D), we see that the model reproduces the assortative arrangement of explorers and exploiters in the network (Pearson correlation coefficient $r\approx0.21$ and no correlation in the rewired case for the reference simulation). This evidence is our second significant result: since the innovation rate $\beta$ is heterogeneously distributed in the synthetic population, and since all agents follow the same evolution rules, the social network's topology influences how individual exploration propensities are distributed. In other words, users feature an increased (decreased) exploration propensity when surrounded by agents with higher (lower) discovery rates.

Furthermore, in Fig.~\ref{fig:simulations}(E) we show that the model correctly reproduces the negative correlation between the average dynamical overlap $o_d$ of an agent with its neighbors and its discovery rate $\beta$ (Pearson correlation coefficient $r\approx-0.12$ and no correlation in the rewired case). As shown in the inset of Fig.~\ref{fig:simulations}(E), the intra-community overlaps ($\langle o_d(C) \rangle_C \approx 0.057$) are also larger than the inter-communities ones ($\langle o_d(C_1,C_2) \rangle_{C_1\neq C_2} \approx 0.004$).
Therefore, the different propensity for an agent to be an explorer (exploiter) depends on its neighbors featuring similar $\beta$ and the opportunity to be exposed to contents and tokens diverse (similar) from those already experienced. In other words, the more varied the tastes between a node and its neighbors are the higher the node's exploration rate. The model pushes users that explore different regions of the content space to easily accept suggestions outside their comfort zone, giving them a higher acceptance probability $P(x,\mathcal{S}_i,c_i,\varepsilon)$. In doing so, users' space of possibilities enlarges, and their $\beta$ exponent is higher. The opposite happens to nodes that explore a limited portion of the contents space. In \SI{8}, we further show that, when we switch off the interaction step of the model, the assortativity and the high overlaps within communities drop significantly.
%
This last evidence is a third tangible effect of a shift from an individual to a collective exploration mechanism. The interaction dynamics between agents can increase their possibility to expand their space of possibilities by receiving suggestions from their neighbors, thus increasing the similarity of listened tokens between friends.

As a final remark, we show in Fig.~\ref{fig:simulations}(F) the  degree distribution $P(k)$ and the community size distribution $P(s)$ obtained letting the social network evolve from an initial all-to-all configuration (with initial weight $w_{ij}=1$ and $\Delta = 0.1$) for about 33.5 millions evolution steps (see the \emph{Materials and Methods} for details on the simulation setup). These distributions feature a scale-free behavior as in the empirical case and confirm that the model also accounts for the emergence of a complex, real-world-like topology based on the interactions between users alone. To summarize, the model reproduces the evolution of the exploration process and the emergence of a complex social network from simple dynamical rules on a semantically structured content space (see \SI{6} for more details).

\section*{Discussion}

The presented empirical study of online music consumption, and the proposed model of collective exploration, highlight the central role of the social environment in shaping how we explore music and discover new content. Our model can indeed reproduce the considerable heterogeneity of individuals' exploration rates empirically observed and the relation between users' rate of discovery and their position in the social network. 
As it turns out, the opportunity to be connected to other individuals with a high propensity to explore makes an individual more likely to explore, on average. Moreover, since all the agents in the model obey the same evolutionary rules, the observed heterogeneity of the individual exploration rates can be explained by stochastic fluctuations (opportunities) and social influence (environment). 
Furthermore, the model reproduces the semantic correlations found in the empirical sequences of music consumption by adequately modulating the individual propensity to select artists similar to their recent history or to move randomly across the whole space of possibilities. It also gives insights into the emergence of communities composed of agents with similar tastes, as shown by the simulations in which the social network is shaped dynamically through random interactions.

%
Some words of caution are in order. 
%
In our analysis, we have supposed that all social links of a user in the social network of \emph{Last.fm} have the same probability of being activated. In real life, one social link can be different from another. Nevertheless, in the absence of knowledge of the actual weights in the empirically studied network, our choice represents, in our view, a good educated guess. A similar issue exists for the explored space of contents, modeled as a static weighted network. Still, in practice, it is continuously evolving in time, shaped by the activity of the users. 
Moreover, our approach did not consider the role played by recommender systems~\cite{gravino2019towards} and other layers of social communications~\cite{battiston2020networks} that may drive users' choices on the platform. 
Finally, we supposed that every user can be modeled with the same evolutionary rules and same set of parameters, while, in reality, different users can naturally have different reinforcement and discovery mechanisms.
We leave the further investigation of these aspects for future work. 

In conclusion, we have focused on a particular type of system: music consumption in online social platforms. We hope our work can represent a significant step forward to develop a general framework to understand how social interactions shape discovery and innovation processes. We reckon that the next important step is to extend our analysis to study exploration processes in economics~\cite{schumpeter1934economic}, biology~\cite{kauffman1993selforganization}, technology~\cite{sole2013innovations, tacchella2020language} and science~\cite{fortunato2018science}. For example, it would be interesting to study scientific collaborations among researchers and their explored topics to get new insights on how success/impact is related to the innovation rate or the position in the social network.


\section{Materials and methods}

\subsection*{Data set}\label{sec:MM_data set}
The \emph{Last.fm} data set presented in this paper contains the complete listening history of a group of 4836 users and their social connections. The users have been selected randomly via a Breadth-First Search algorithm starting from a random seed node, using \emph{Last.fm}'s API method \texttt{user.getFriends()}.
More in detail, at each step, a random user is chosen from the neighboring nodes of those users selected up to that point, and at least 10'000 listening records. Then, we download the complete history of streamed tracks of the selected user through the API \texttt{user.getrecenttracks()} endpoint. For each record, besides the timestamp, the data set provides additional metadata, including the name and MusicBrainz Identifier (MBID) of the track, artist, and album if present. In total, the records span over almost 13 years, from August 2005 to February 2018, with 335'375'125 unique streaming events, totaling 6'972'047 tracks, 958'732 artists, and 1'807'150 albums. In total, 31.8\% of the set of tracks does not have an album, making up for 9.1\% of all listening records in the data set. 
Considering the vast number of different tracks and the lack of consistency of albums, the analysis in this paper has been carried out focusing only on the sequences of artists that a user listens to---even though similar results can be obtained when considering the sequence of tracks instead.
The data set, collected in this way, is unique in nature and breadth. The music listening histories data set presented in Ref.~\cite{vigliensoni2017music} consists of more than 27B logs from 583k users, for a total of 555k different artists and 46k logs per user on average. Still, no social relations between the users are given, similarly to Refs.~\cite{lastfm360k1k,brost2019music}. On the contrary, in Ref.~\cite{Cantador:RecSys2011} the social relationships are present, but for each user there is only the tag assignments history, unfortunately much shorter (around 98 tracks on average).
The dataset used to reproduce the results is available for download on figshare \url{http://dx.doi.org/10.6084/m9.figshare.16652104}. The code used to reproduce the results is available upon request by contacting the corresponding author.

\subsection*{Entropy measure} To measure the presence of semantic correlations in the listening sequences, we compute the Shannon entropy of the $f_i^a$ occurrences of each artist $a$ in the sequence $\mathcal{S}_i$ (whose length is $T_i$) of user $i$, as in Ref.~\cite{tria2014dynamics}. We identify the sub-sequence $\mathcal{S}_i^a$ of $\mathcal{S}_i$ starting at the first occurrence of $a$. We then divide $\mathcal{S}_i^a$ in $f_i^a$ parts of equal length and count the number $f_i^a(x)$ of occurrences of $a$ in each $x^\text{th}$ interval. Then, the normalized Shannon entropy of artist $a$ in $\mathcal{S}_i$ is defined as $S_i^a(f_i^a) = -\frac{1}{\log f_i^a} \sum_{x=1}^{f_i^a} \tilde f_i^a(x) \log{\tilde f_i^a(x)}$, where $\tilde f_i^a(x) = {f_i^a(x)}/{f_i^a}$.
When the occurrences are equally distributed in the $f_i^a$ intervals, $S_i^a(f_i^a)$ hits its maximal value $1$, whilst when all the events are found in a single interval the entropy is at its minimum value $S_i^a(f_i^a)=0$. Finally, the average Shannon entropy is computed as $S(f) = \langle S_i^a(f_i^a) \rangle_{i,a | f_i^a = f}$, where we define the weight $w_f$ as the number of times $|\{i,a | f_i^a = f\}|$ an artist $a$ appears in a sequence $\mathcal{S}_i$ with frequency $f$.
Therefore, low values of $S(f)$ and high values of $\langle\Delta S \rangle$ are related to the presence of non-trivial semantic correlations in the process.



\subsection*{Structure of the network of contents}
We extract the semantic structure of the network of artists starting from the reduced sequence $\tilde{\mathcal{S}_i}$ of each user $i$, i.e., the temporally-ordered sequence of artists listened by $i$ without consecutive repetitions (multiple consecutive streams of one artist reduce to a single event). We then measure the number of times $N_{a \to b}$ users listen to an artist $b$ after having listened to artist $a$, with $a\neq b$ as in Fig.~\ref{fig:artists_network}(A). Since we aim at an undirected network of proximity, we define $N_{ab} = N_{a \to b} + N_{b \to a}$. We then shuffle $Q = 100$ times each sequence $\tilde{\mathcal{S}_i}$, counting for each realization $q$ the number $N^q_{ab}$. We then compute the expectation $\overline N_{ab} = \langle N^q_{ab}\rangle_q$ and the standard deviation $\sigma_{ab} = \rm{std}(N^q_{ab})$ of the $N_{ab}$ count under the assumption that there are no semantic relations in the sequences of streams. We hence define the proximity of two artists via the z-score $z_{ab} = (N_{ab} - \overline N_{ab}) / \sigma_{ab}$, obtaining a proximity network between the artists where a link is drawn if two artists $a$ and $b$ feature a z-score $z_{ab} > \bar z$, where $\bar z$ is a fixed threshold. Finally, we disregard all artists with overall number of appearances in the reduced sequences less than a threshold $\bar f$. Using $\bar z = 1$ and $\bar f = 100$ we obtain an undirected, weighted network $\mathcal{G}_C$ with 266'694 nodes and 17'765'819 edges, where we set the weight $w_{ab}$ of a link to be $w_{ab} = min(z_{ab},100)$, directly related to the closeness of the two artists.

\subsection*{Sampling of semantic and novelties tokens}
During the exploration step of the model, for the extraction from the active agent $i$'s urn $\mathcal{U}_i(t)$, different items (artists) have different probabilities of being drawn. Each item $a$ is indeed present in $\mathcal{U}_i(t)$ with a multiplicity $w_i^a(t)$ of identical balls, added in previous reinforcement steps or through social recommendations. Besides this, we temporarily reduce the probability of jumping to nodes semantically distant from $i$'s recent history in order to increase semantic correlations in the sequences. 
In particular, $2(\nu+1)$ nodes $a$ are randomly sampled from the neighbors---in $i$'s space of possibilities $\mathcal{G}_C^i(t)$---of the last visited node, keeping the original weight $w_i^a(t)$, as well as up to $2(\nu+1)$ nodes surrounding the last node accepted via social interaction if this was successful during $i$'s last step. The weight of all other nodes are temporarily reduced by a factor $\eta \leq 1$, i.e., $w_i^a(t) \to \eta w_i^a(t)$. 
For $\eta=1$, this individual exploration process is equivalent to the urn model with triggering without semantics~\cite{tria2014dynamics}, while for $\eta = 0$ this corresponds to a standard random walk with node reinforcement on the content space (without jumps). 
Moreover, if the chosen destination $x$ is new for the agent $i$, i.e.,  $x \notin \mathcal{S}_i(t)$, the discovery event triggers the insertion of novelties in $i$'s urn, including $\nu+1$ new nodes randomly selected from the neighbors $a$ of $x$ using the weights $z_{xa}$ in $\mathcal{G}_C$.
In this case, if the number of new neighbor nodes---not already present in $i$'s space of possibilities $\mathcal{G}_C^i(t)$---is less than $\nu+1$, the search extends to nodes at a distance of two from $x$. 

\subsection*{Acceptance probability estimation}
When an agent $i$ evaluates a token $x$ proposed by one of its neighbors, $i$ accepts it with probability $P=P(x, \mathcal{S}_i(t),c_i,\varepsilon)$, where $c_i$ represents the core of $i$, that is the set of the $c$ most frequent nodes visited by $i$. We define the distance of a node from the core as $d_i^c(x)=\min_{y\in c_i}\left\{\text{dist}(x,y)\right\}$. Then, in order to calculate $P$, we compute the following density function: 
\begin{equation}
    p(d, \mathcal{S}_i(t),c_i) = \frac{\left|\left\{ x\in\mathcal{S}_i(t) : d_i^c(x) = d\right\}\right|} {\left|\mathcal{S}_i(t)\right|}.
\end{equation}
Therefore, we can calculate $P$ as
\begin{equation}
    P(x, \mathcal{S}_i(t),c_i,\varepsilon) = \sum_{d=0}^{d_i^c(x)} p(d, \mathcal{S}_i(t),c_i) \pm \varepsilon.
\end{equation}
No significant difference in the analysed observables has been detected changing $c$ and $\varepsilon$, that we have hence fixed to $c=10$ and $\varepsilon=0.1$.

\subsection*{Selection of parameters}
To select the reference set of parameters, we first consider for which values of $\eta$ it is possible to obtain a Shannon entropy weighted difference $\langle\Delta S \rangle$ similar to the empirical one. As shown in \SI{5}, this can only happen for $\eta\leq 0.01$. Therefore, we constrain the search to simulations with $\eta\leq 0.01$.
Then, we simulate each set of parameters $\mathbf{x}$ and we compute the Kullback-Leibler divergence  $KL(\mathbf{x})$ between the empirical $P(\beta)$ and the simulated one.
%
%
We hence select the reference set of parameters $\mathbf{x}$ such that $KL(\mathbf{x})$ is minimum. 
Since the simulations are heavily time- and computing-demanding, we explore the parameter space limiting ourselves to the 10\% of the original sequences in the data set, that is about 33.5 million total steps. In \SI{7} we show how the reference set of parameters found in the limited sequence matches the empirical findings even when we let the simulation run for as many steps as the empirical number of streams.

\subsection*{Dynamical network simulations}
In the case of a dynamically evolving social network, we start from an all-to-all network $\mathcal{G}_S$, giving weight $w_0 = 1$ to all the edges between all the pairs of users. 
During each interaction step, the neighbor $j$ to which the active agent $i$ asks for a suggestion is selected with probability $p(j) = w_{ij} / \sum_{k\sim i} w_{ik}$. Each time $i$ interacts with $j$ we let the respective edge weight evolve according to the law $w_{ij}(t+1) = w_{ij}(t) \pm \Delta$, with $\Delta=0.1$ and with a sign + (-) when the interaction between $i$ and $j$ at time $t$ is positive (negative). During the simulation, we clip each edge weight such that $0.1 \le w_{ij} \le 10$. To better characterize the agents in the simulation, we let the first 1\% of each individual's events to be equal to their original sequence as a warm-up (see~\SI{6} for details).

\section*{Acknowledgements}
I.I. acknowledges support from the James S. McDonnell Foundation $21^{\text{st}}$ Century Science Initiative Understanding Dynamic and Multi-scale Systems - Postdoctoral Fellowship Award, from the Agence Nationale de la Recherche (ANR) project DATAREDUX (ANR-19-CE46-0008), from the United Kingdom Regions Digital Research Facility (RDRF)-Urban Dynamics Lab under the Engineering and Physical Sciences Research Council (EPSRC) Grant No. EP/M023583/1 and from The Alan Turing Institute under EPSRC Grant No. EP/N510129/1. 
V. Latora acknowledges support from the Leverhulme Trust Research Fellowship RF-2019-059 “CREATE: the network components of creativity and success”.


\section*{Competing interests}
The authors declare no conflict of interest.

\section*{Author contribution statement}
GDB, EU, BM, and V Loreto designed research; GDB and BM crawled and filtered the data set; GDB and EU performed research and analysis; all authors contributed to develop the model and to write the paper.

\bibliography{arxiv}


\clearpage
\setcounter{figure}{0}
\setcounter{table}{0}
\setcounter{equation}{0}
\makeatletter
\renewcommand{\thefigure}{S\arabic{figure}}
\renewcommand{\theequation}{S\arabic{equation}}
\renewcommand{\thetable}{S\arabic{table}}

\setcounter{secnumdepth}{2} 
\renewcommand{\thesection}{\arabic{section}} 
\renewcommand{\thesubsection}{\thesection.\arabic{subsection}}

\widetext
\begin{center}
	\textbf{\large Supplemental Information: Social interactions affect discovery processes.}
\end{center}

 
\section{Social network sample statistics}

\begin{figure}[!ht]
    \centering
    \includegraphics[width=.5\textwidth]{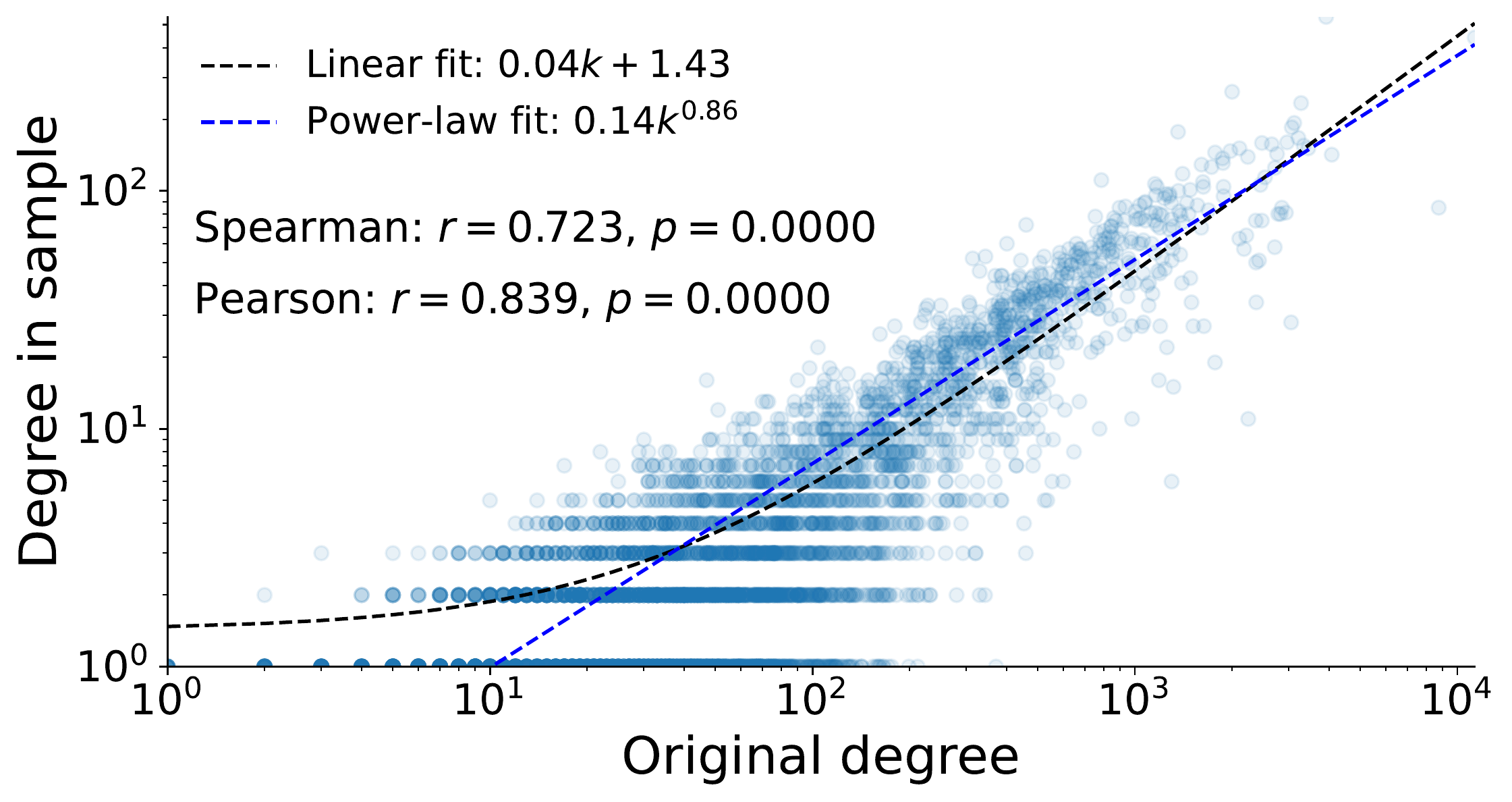}
    \caption{We show the in-sample  degree (y axis) versus the original number of friends the \textit{Last.fm} dataset for all the users. Spearman and Pearson correlation coefficient are shown in the legend. We also show a linear $a k + q$ (black dashed line) fit and a power-law $ak^b$ (blue dashed line) one.}
    \label{fig:s01_degree}
\end{figure}

We have crawled the \emph{Last.fm} platform growing a breadth-first search sub-graph from a random seed, ending up with a connected network made of 4836 users. 
In Fig.~\ref{fig:s01_degree} we show that this sample significantly reproduces the degree that users feature on the \textit{Last.fm} platform, with a Spearman rank correlation $r=0.723$. ($p<0.0001$) and Pearson $r=0.839$. ($p<0.0001$) between the degree of a user in the sample network and the total number of followed users on the platform.
In particular, we observe that the degree in the sample is, on average, almost one tenth of the original one, thus providing us with a subset of the original connections of each user.

\section{Estimation of the Heaps' exponents}
\begin{figure}[!ht]
    \centering
    \includegraphics[width=.5\textwidth]{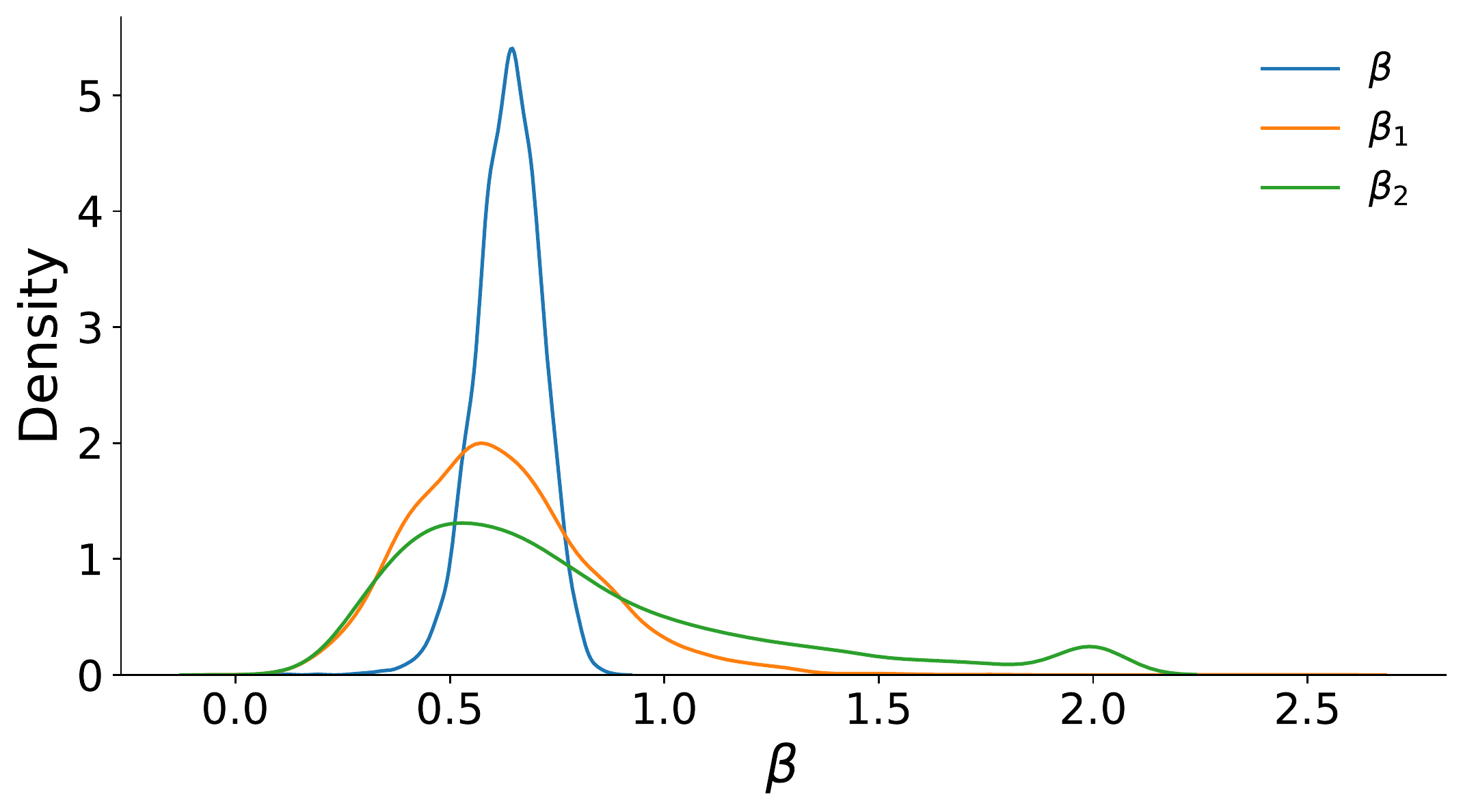}
    \caption{Comparison of the Heaps' exponents distribution on the sequences in the dataset according to three different approximation methods.}
    \label{fig:s02_approximation}
\end{figure}

In this work, we have estimated the Heaps' exponent as the slope in the log-log scale of the function $D(t)$ counting the number of different tokens present in the sequence up to length $t$. In particular, if the sequence has length $T$, it is calculated as $ \beta = \log(D(T)) / \log(T)$. On average, for a power law behavior, this method gives a good approximation of the overall slope. However, sometimes the slope of the power law can change over time, and one might be interested in the slope of the tail specifically. This way, though, there might be cases in which the approximated slope is higher than 1, which is not theoretically possible for the Heaps' law, since $D(t)\leq t$ is hardly constrained to be less than the linear function. This is the same reason why we have not used the other approximations where, given the points ($t$, $D(t)$), the reference parameters that approximate $D(t)$ are found as a function $a x^{\beta_1}$ or $(1 + x/a)^{\beta_2}$, where $\beta_1$ and ${\beta_2}$ would be the approximated exponents. The distributions of the Heaps' exponents according to these three approximations is shown in Fig.~\ref{fig:s02_approximation}. As it can be seen, the number of cases for which $\beta_1$ or $\beta_2>1$ is significant. In all cases, it is worth saying that calculating the Heaps' exponent with these approximations at 10\% of the sequence gives a different result than doing it for the whole sequence, showing that these are not perfect power laws (see \textcolor{blue}{{\it SI Appendix} 8} for more details).

\section{Zipf's and Heaps' exponent correspondence}
\begin{figure}[!ht]
    \centering
    \includegraphics[width=.5\textwidth]{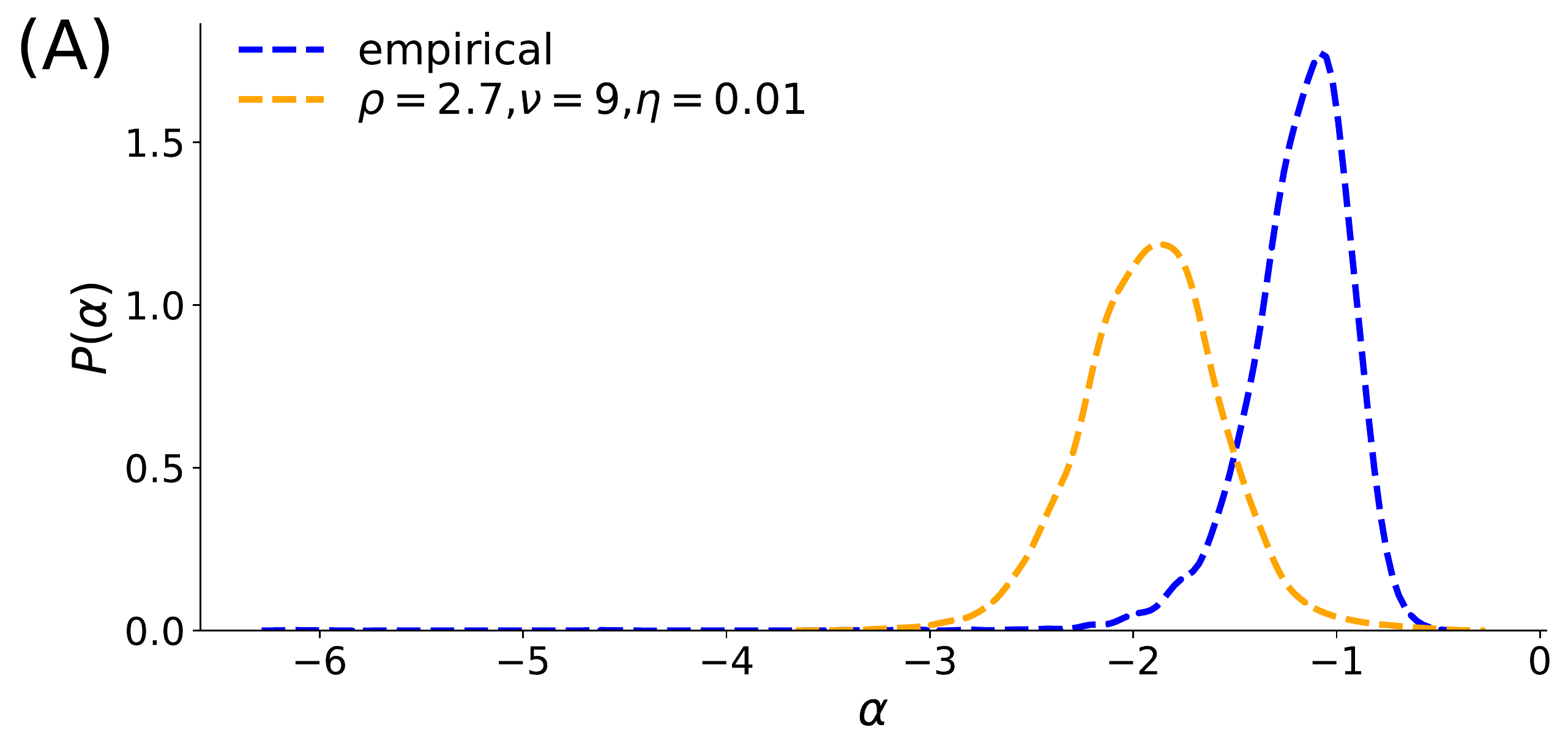}\includegraphics[width=.5\textwidth]{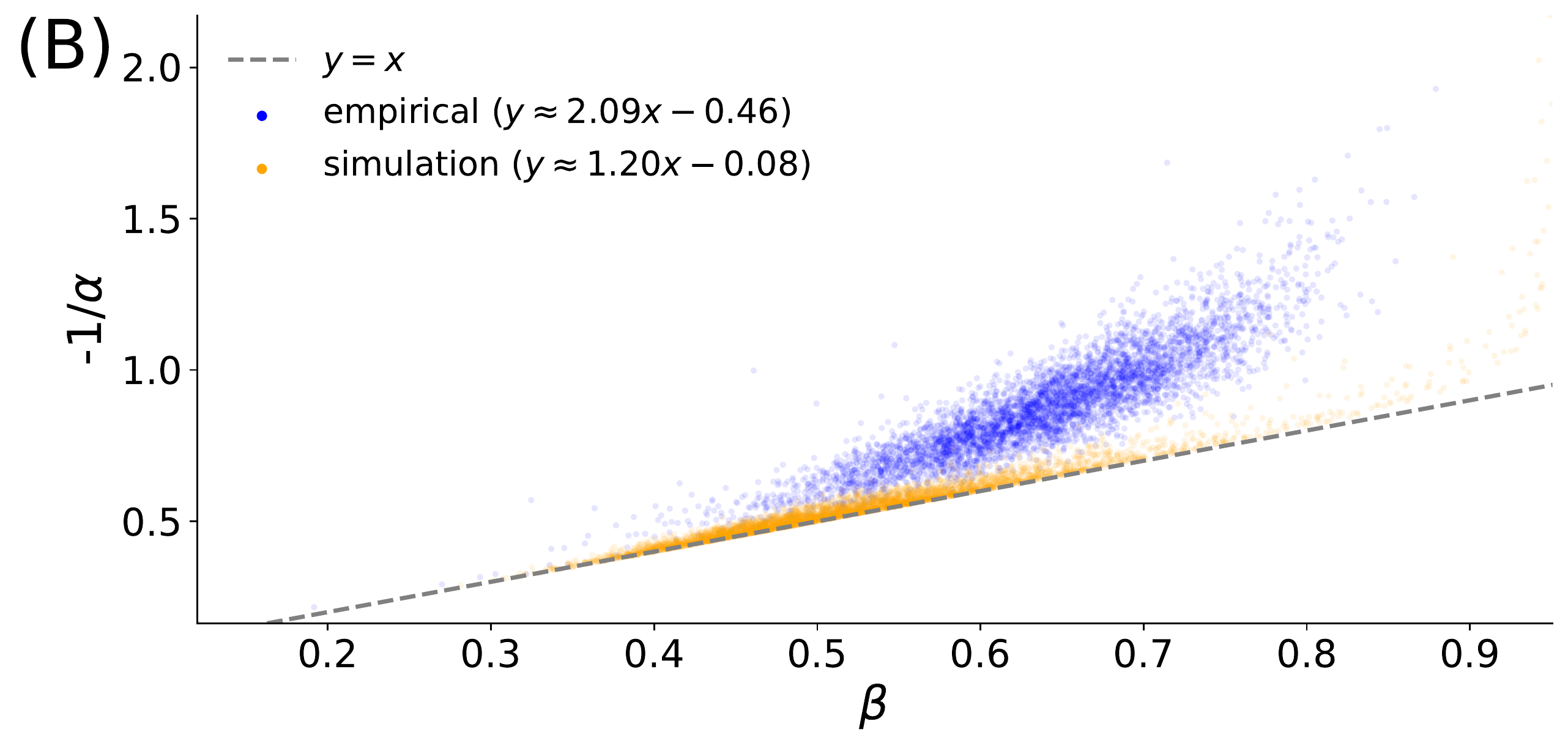}
    \caption{\textbf{(A)} Comparison of the Zipf's exponents distribution on the sequences in the dataset (blue) and of the reference simulation (orange). 
    \textbf{(B)} Scatter plot between the Heaps' exponents and the anti-reciprocal of the  Zipf's exponents for the sequences in the dataset (blue) and of the reference simulation (orange), with the respective linear regression displayed in legend.}
    \label{fig:s03_zipf}
\end{figure}

In the main text we have discussed at plenty about the Heaps' law and its significance. The Zipf's law has also been of great importance~\cite{Zipf}.  
Decreasingly ordering the occurrence frequencies $f$, the Zipf's law can be expressed as $f(R) \sim R^{-\alpha}$, where $R$ indicates the rank and $\alpha$ is called \textit{Zipf's exponent}. These two laws have been observed in various empirical systems, producing different values of $\alpha$ and $\beta$~\cite{cattuto2007semiotic,cattuto2007vocabulary,tria2014dynamics,monechi2017waves}, and under mild assumptions they are asymptotically equivalent~\cite{serrano2009modeling, lu2010zipf, gerlach2013stochastic}, being one the anti-reciprocal of the other. For this reason, all of the models that have been recently proposed keep these laws into considerations, trying to reproduce the innovation distributions observed in the empirical datasets~\cite{loreto2016dynamics}. 
In Fig.~\ref{fig:s03_zipf}(A) we show the distributions of the Zipf's law for the sequences in the empirical dataset (blue) and the reference simulation found in our analysis (orange). Moreover, in Fig.~\ref{fig:s03_zipf}(B) we compare the Heaps' exponents $\beta$ and the anti-reciprocal $-1/\alpha$ of the  Zipf's exponents. As analytically found for the UMST~\cite{tria2014dynamics}, in the simulation we observe an almost linear correlation between the two exponents (Pearson $r=0.94$, $p<0.001$), although in the empirical case it is less strong (Pearson $r=0.90$, $p<0.001$), with a linear regression $-1/\alpha \approx 1.20 \beta - 0.08$ in the simulation against $-1/\alpha \approx 2.09 \beta - 0.46$ in the data.

\section{Agent's urn initialization procedures}
At the beginning of the simulation, each agent's urn is initialized with some balls representing tokens of the semantic space, so that the exploration of such space starts from these points. 
The problem of the initialization is related to whether to choose more widely distributed initial nodes, or, on the other extreme, the same node for every agent, and how this choice can affect the results.
It turns out that this choice has little impact on the outcome of the simulations, only delaying or reducing some effects due to interactions with peers initially too distant or too close. 
To solve this problem, since in this work we have considered the same agents as in the Last.fm data set crawled, 
in our simulations we initialize the agents urn with the first 
artists they listened to in the empirical data. This solution is natural when introducing an initial warm-up phase, in which the first steps of the model are not randomly chosen according to the rules of the model, but are instead the same steps made by the corresponding user in the empirical data. The impact of the length of the warm-up phase is shown in \textcolor{blue}{{\it SI Appendix} 6}. Finally, 
we want to notice that another viable option for the initialization of the urns could have been to choose the $\nu+1$ most frequent artists in their own listening history.

\section{Selection of the reference set of parameters}
In our analysis of the model, we have first developed a method to select a reference set of parameters that reproduce the main characteristics present in the data. We numerically simulate the model using the fixed social network from Last.fm and the space of tokens given by the artists listened in the dataset (see main text). Moreover, for this search of the reference set of parameters, we run the simulations for only the first 10\% of the real sequences due to computational reasons, for a total of around 33.5 millions evolutionary steps.
We explore the space of parameters varying the set of parameters ($\eta$, $\rho$, $\nu$). 
In particular, we try the following values for the jump probability parameter $\eta$: $0.001$, 0.01 0.05, 0.1, 0.2, 0.3, \dots , 1.
Given $\eta$, the reinforcement parameter $\rho$ spans from 1 to 4 at steps of 0.1. Then, based on the results given in \cite{tria2014dynamics} on the limits of the Heaps' exponents in the UMST, we consider integer values of the triggering parameter $\nu$ ranging from 0 to $\max(20,\,1.5\rho/\eta)$. No significant difference has been observed changing $c$ or $\varepsilon$, that we hence fix to $c=10$ and $\varepsilon=0.1$. 

\begin{figure}[!ht]
    \centering
    \includegraphics[width=.5\textwidth]{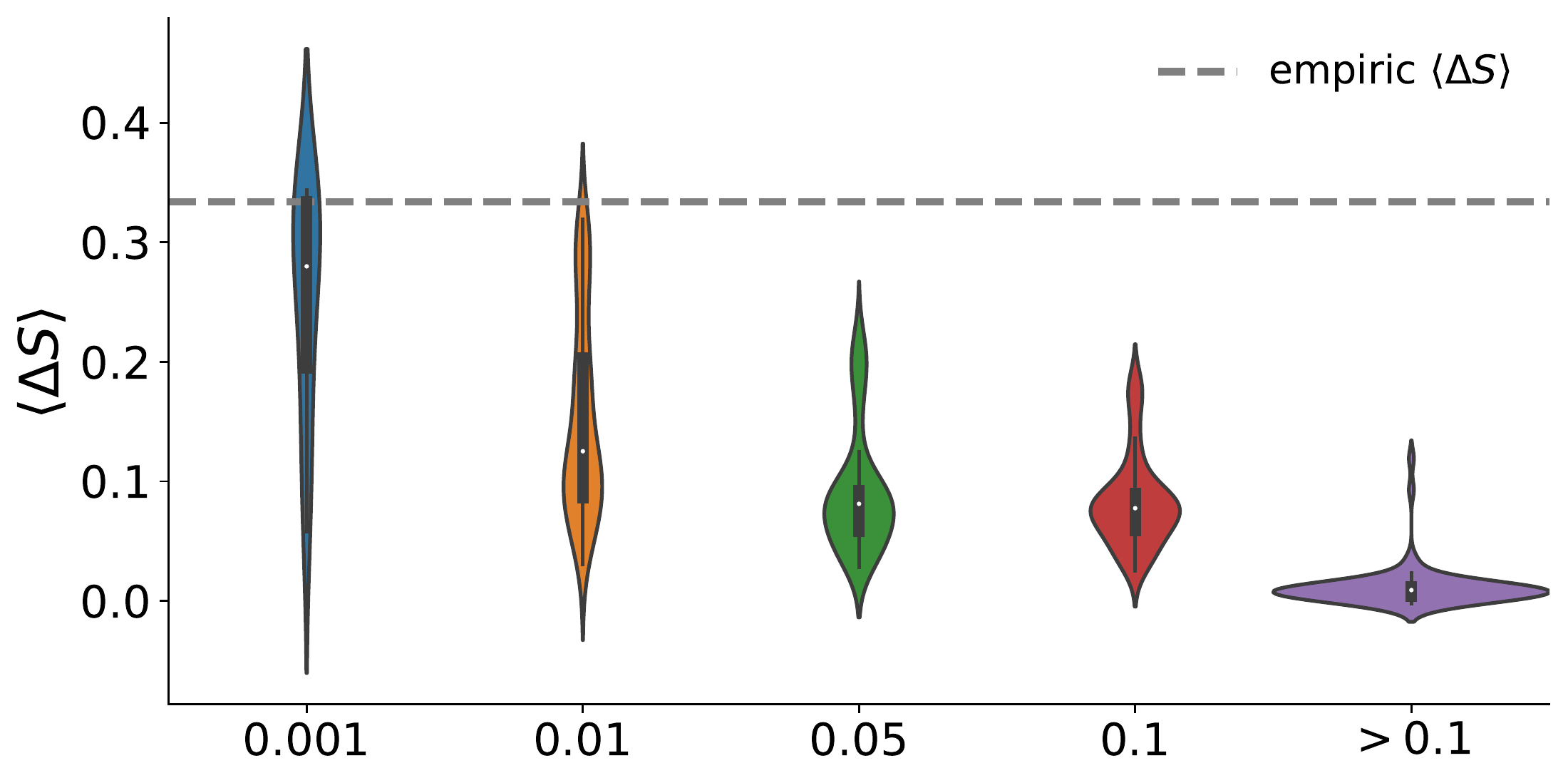}
    \caption{Violin plot showing the distributions of the Shannon Entropy weighted difference $\langle\Delta S\rangle$ with the randomized sequences, grouping the various simulations together according to their value of $\eta$. Notice that only the distributions relative to $\eta=0.001$ and $\eta=0.01$ overlap the value of $\langle\Delta S\rangle = 0.334$ of the empiric sequences stopped at 10\%, i.e. same length of the simulations.}
    \label{fig:s06_reference_simulations}
\end{figure}

We focus on just two basic footprints of discovery processes, namely the Heaps' exponent distribution, responsible for the balance between exploit and explore strategies, and the normalized Shannon Entropy of the distributions of artists in the listening sequences, which captures the presence of semantic correlations during music consumption. 
Since these two measures are not comparable, we first find the reference value of $\eta$ by considering, for each simulated set of parameters $\mathbf{x}$, the normalized Shannon entropy weighted difference $\langle\Delta S^{(\mathbf{x})}\rangle =  { \sum_{f} w_f (\tilde{S}^{(\mathbf{x})}(f)-S^{(\mathbf{x})}(f))}/{\sum_f w_f}$ between the normalized Shannon entropy $S^{(\mathbf{x})}(f)$ calculated on the sequences generated by the simulation and the one $\tilde{S}^{(\mathbf{x})}(f)$ for the reshuffled ones. 
We group the various values of $\langle\Delta S^{(\mathbf{x})}\rangle$  by the parameter $\eta$ of the corresponding simulation, that is $0.001$, 0.01 0.05, 0.1, and $>0.1$ (0.2, 0.3, \dots , 1), plotted in Fig.~\ref{fig:s06_reference_simulations}. 
We notice that only the groups of $\eta=0.01$ and $\eta=0.001$ range in an interval that includes the value of $\langle\Delta S\rangle$ calculated in the empiric data  set in their first 10\%. 
Therefore, we restrict our search only to simulations with $\eta \leq 0.01$. 
Then, for each simulated set of parameters  $\mathbf{x}$, we calculate how much the Heaps' exponents distribution differs from the real one. To do this, we compute the 
%
%
Kullback-Leibler divergence  $KL(\mathbf{x})$ between the empirical $P(\beta)$ and the synthetic one.
Finally, we select the reference set of parameters $\mathbf{x}$ as the one among those with $\eta\leq 0.01$ such that $KL(\mathbf{x})$ is minimum.
The first six sets of parameters given by this process are given in Table~\ref{table:reference_simulations}, together with the values of the Shannon entropy weighted difference $\langle\Delta S(\mathbf{x})\rangle$ and the Kullback-Leibler divergence  $KL(\mathbf{x})$.
We don't find particular differences between simulations with the same quotient $\rho/\nu$ and same $\eta$, so that all reference simulations have a quotient value between 3 and 3.5. 

\begin{table}[!ht]
\begin{center}
\begin{tabular}{cccccc}
\toprule
{} &  $\rho$ &  $\nu$ &   $\eta$ &     $KL$ &  $\Delta S$ \\
\midrule
\textit{1.}  &  2.7 &   9 &  0.01 &  0.213398 &    0.239159 \\
\textit{2.}  &  3.0 &  10 &  0.01 &  0.216276 &    0.232123 \\
\textit{3.}  &  3.1 &  10 &  0.01 &  0.216367 &    0.283424 \\
\textit{4.}   &  2.0 &   7 &  0.01 &  0.218269 &    0.207541 \\
\textit{5.}  &  2.1 &   7 &  0.01 &  0.220671 &    0.243933 \\
\textit{6.}   &  2.9 &  10 &  0.01 &  0.222252 &    0.239109 \\
\bottomrule
\end{tabular}
\end{center}
\caption{The first six sets of parameters according to the $KL$ sorting, with $\eta \le 0.01$. 
}
\label{table:reference_simulations}
\end{table}


\section{Dynamical network simulations}\label{sec:dynamical}
\begin{figure}[!bt]
    \centering
    \includegraphics[width=.45\textwidth]{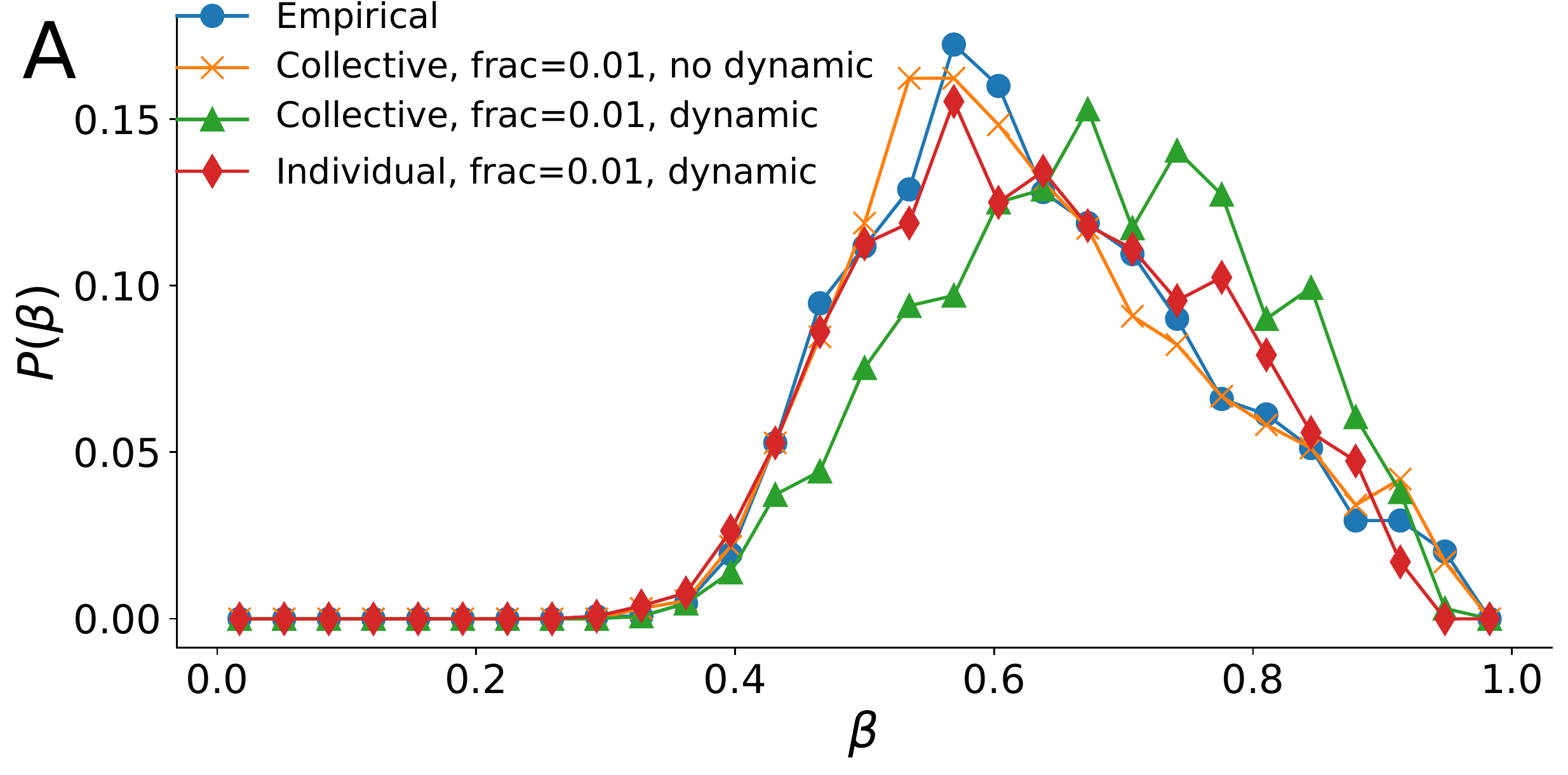}
    \includegraphics[width=.45\textwidth]{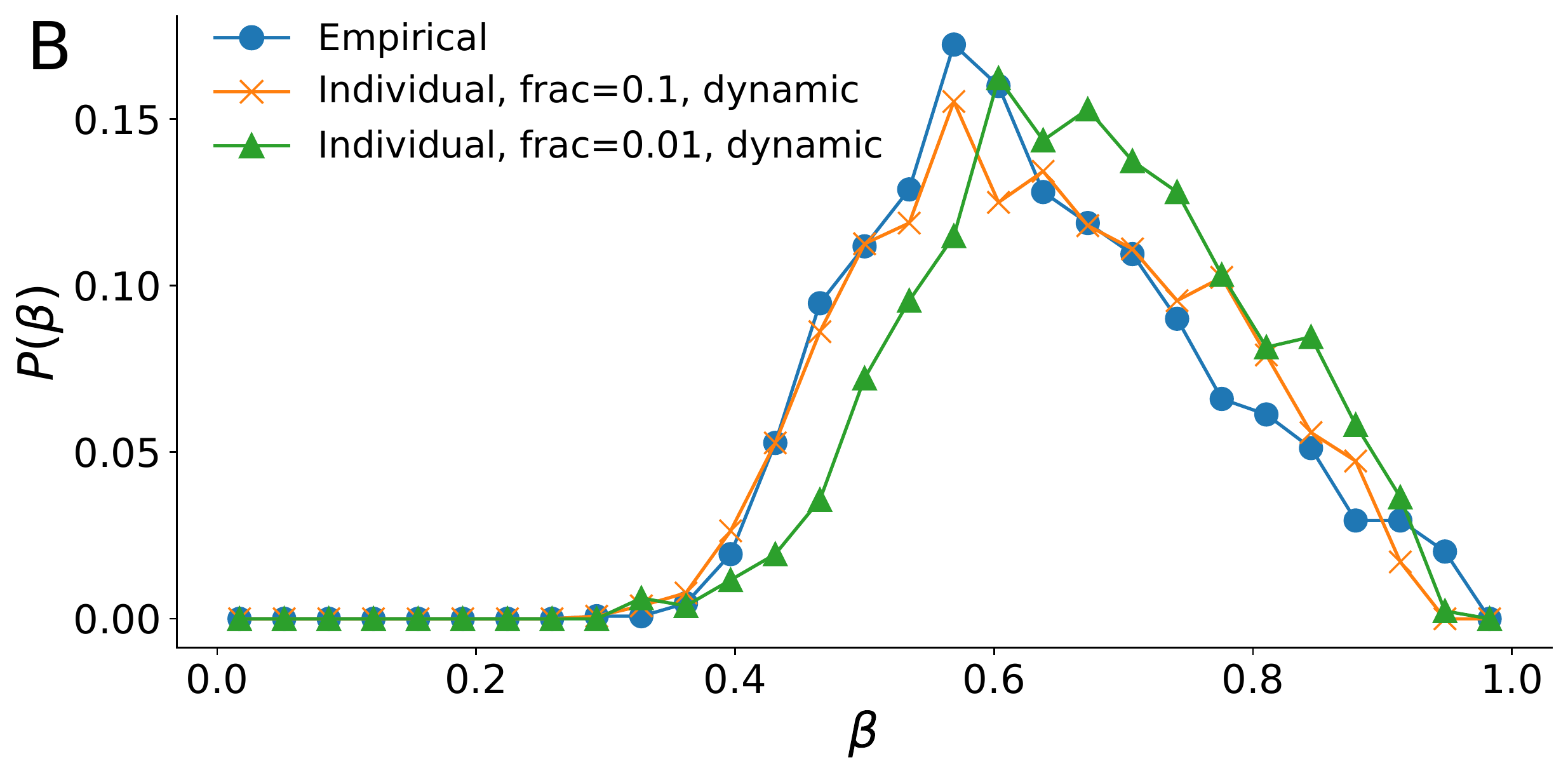}
    \caption{
    \textbf{(A)} Comparison of the $P(\beta)$ distribution in the empirical case (blue line) and in the simulations with the reference set of parameters, using a collective warm-up for 1\% of the events with (orange line) and without (green line) social dynamics on the edges and with an individual warm-up for 1\% of the events with edge dynamics (red line).
    \textbf{(B)} The same but limited to an individual warm-up of 10\% of the events (orange line) and 1\% (green line).}
    \label{fig:s06_pbeta_warm-up}
\end{figure}

Here, we show that our results are robust with respect to the change of warm-up strategy and duration as well as the presence, or not, of social edge weight changes after each interaction. 
We use either individual or collective warm-up for a certain fraction of events, meaning that the first fraction of, respectively, the agent's or the population events in the simulations corresponds exactly to the one in the empirical listening sequences.
In Fig.~\ref{fig:s06_pbeta_warm-up}A we show that the $P(\beta)$ distribution found in the empirical case is still well reproduced if we change the kind of warm-up. Additionally, in Fig.~\ref{fig:s06_pbeta_warm-up}B we show that the model reproduces the original distribution also varying the warm-up duration and keeping fixed the other parameters to the reference set of parameters.

\begin{figure}[ht]
    \centering
    \includegraphics[width=.45\textwidth]{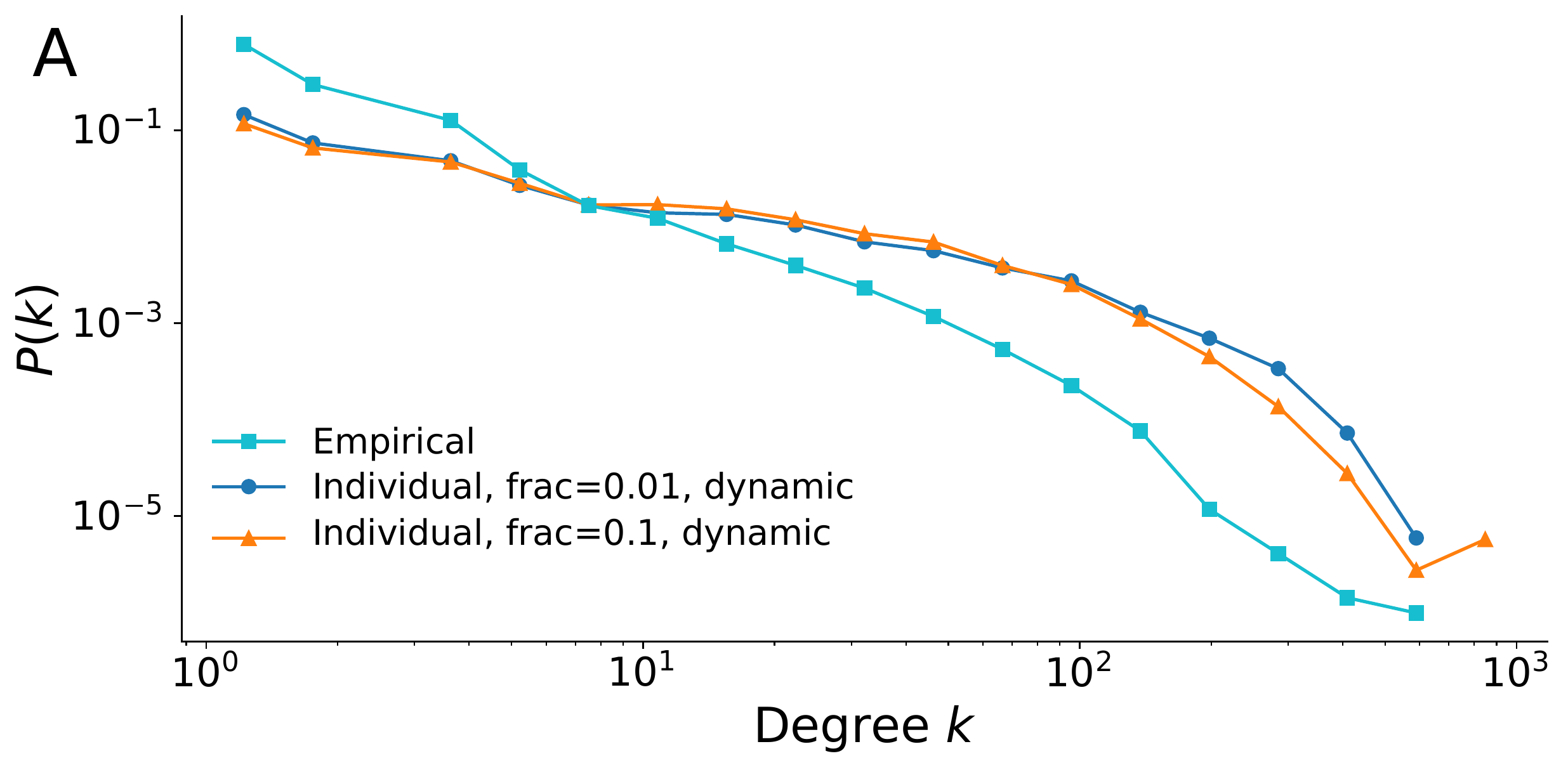}
    \includegraphics[width=.45\textwidth]{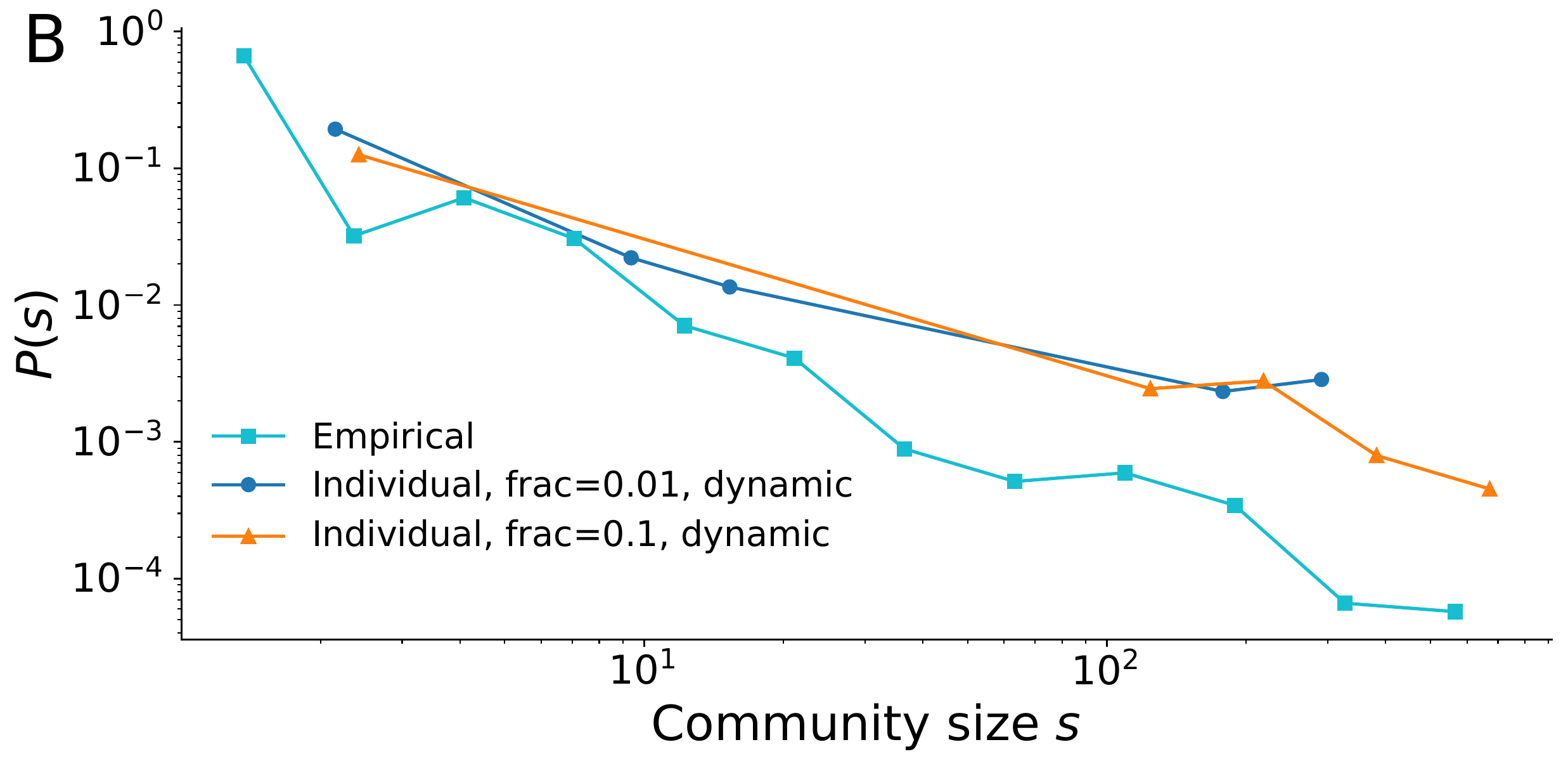}
    \caption{
    \textbf{(A)} Comparison of the $P(k)$ degree distribution using $\rho=2.7,\,\nu=9,\,\eta=0.01,\,c=10$ and $\varepsilon=0.1$ in the empirical case (cyan line), and using an individual warm-up for 1\% of the events (blue line) and for the 10\% of them (orange line) with dynamics on the edges.
    \textbf{(B)} The same but focusing on the $P(s)$ communities size distribution.}
    \label{fig:s07_degCommDistrib_warm-up}
\end{figure}

The results are robust with respect to the change of warm-up also when inspecting the degree distribution $P(k)$ (see Fig.~\ref{fig:s07_degCommDistrib_warm-up}A) and community size distribution $P(s)$ (see Fig.~\ref{fig:s07_degCommDistrib_warm-up}B).
Notably, in the simulations where we switch on the dynamics on the edges, we start from an all-to-all social network with initial weight $w_0 = 1$ and increment equal to $\Delta = 0.1$ after every interaction, clipping the values between $w_{min} = 0.1$ and $w_{max} = 10$. With such rules, the model is able to reproduce both the degree and community size distribution, as well as the $P(\beta)$ distribution, highlighting the goodness of the chosen modeling framework to reproduce the empirical findings. 
Note that, in the simulations with the edge dynamics turned on, we draw an edge between nodes $i$ and $j$ if the edge's weight $w_{ij} > \bar w$, where $\bar w$ is the largest weight cut-off at which we have a single weakly connected component in the directed social network.

\begin{figure}[ht]
    \centering
    \includegraphics[width=.5\textwidth]{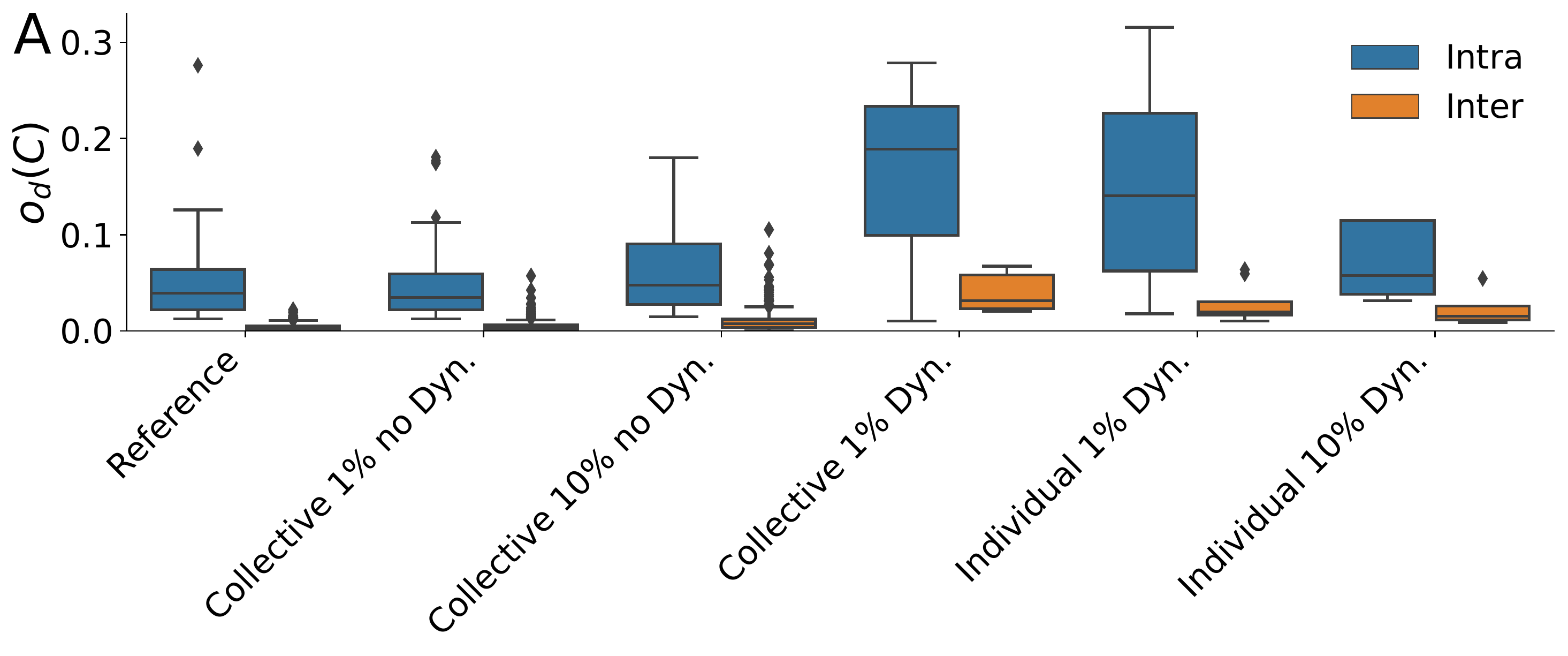}\\
    \includegraphics[width=.5\textwidth]{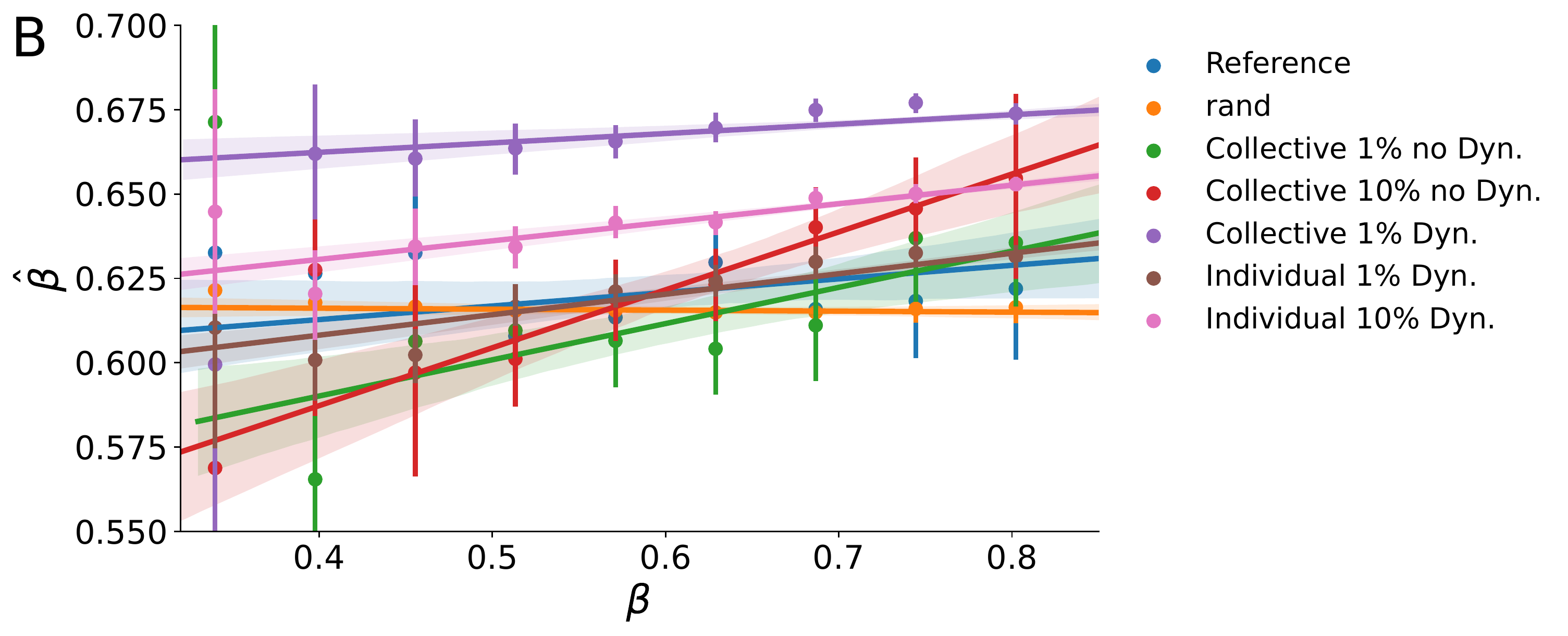}%
    \includegraphics[width=.5\textwidth]{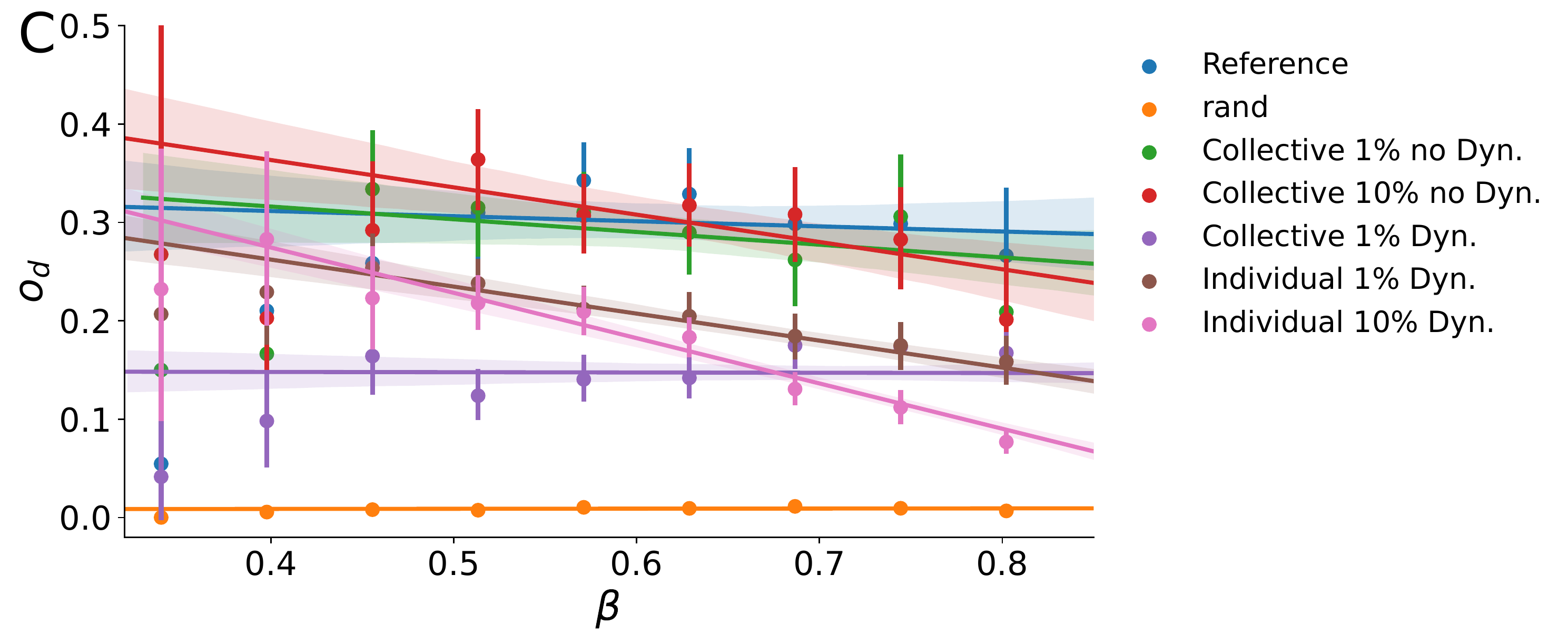}
    \caption{
    \textbf{(A)} Comparison of the $P(o_d(C_i,C_j))$ distribution of the average dynamical overlap found within a community (blue box, intra) and between (orange box, inter) two distinct communities for different simulations settings (see x axis labels for details). 
    \textbf{(B)} The assortativity analysis between the node $\beta$ and the $\hat \beta$ average of its neighbors. Different warm-up and edge dynamics strategies are reported in the legend.
    \textbf{(C)} The same analysis but focusing on the assortativity of $\beta$ with the dynamical overlap with neighbors $o_d$. 
    In all the panels the \textit{Reference} case corresponds to the reference parameters set $\rho=2.7,\, \nu=9,\, \eta=0.01$. 
    The slope of the correlations found are reported in Table \ref{table:fit_warm-up}.
    }
    \label{fig:s08_over_assort_warm-up}
\end{figure}

Remarkably, the model reproduces the empirical dynamical overlap $o_d$ and the assortativity between a node's Heaps' exponent $\beta$ and its neighbors' one $\hat{\beta}$ also with the dynamics on the social edges. Moreover, the assortativity between the focus node's $\beta$ and its dynamical overlap $o_d$ with neighbors is robust with respect to the change of the warm-up kind and duration as well as to the presence of a dynamics on the social network's edges weight.
As we show in Fig.~\ref{fig:s08_over_assort_warm-up}(B,C) and Table~\ref{table:fit_warm-up}, the individual warm-up with edge dynamics turned on results in a higher positive assortativity of $\beta$ and $\hat\beta$ as well as in a stronger negative correlation between $\beta$ and $o_d$, more in line with the empirical findings. This is because the individual warm-up allows to better characterize the tastes of a single node, by imposing a larger fraction of its first exploration events.
Moreover, we see in Fig.~\ref{fig:s08_over_assort_warm-up}(A) that the edge dynamics, as expected, drives the system to communities with a higher internal overlap with respect to the non dynamical case. Notably, if we let the nodes to be characterized enough (individual warm-up with the first 10\% of the node's events fixed to the original sequence, see Figure \ref{fig:s08_over_assort_warm-up}(A)), the intra- and inter-communities overlap value approaches the empirical one (intra-community $\langle o_d(C)\rangle_C \sim 0.07$ inter-community $\langle o_d(C1,C2)\rangle_{C_1\neq C2} \sim 0.025$).

\begin{table}[!ht]
\begin{center}
\begin{tabular}{cccccc}
\toprule
Warmup &  Kind &  Fraction &   Edge dynamics &  $r(\hat\beta)$ &  $r(o_d)$ \\
\midrule
No, reference case  &  n.a          &  0    & No    &  0.05     &    -0.018 \\
No, reshuffled case &  n.a          &  0    & No    &  -0.05    &    0.003 \\
Yes                 &  Collective   &  1\%  & No    &  0.12     &    -0.048 \\
Yes                 &  Collective   & 10\%  & No    &  0.18     &    -0.095 \\
Yes                 &  Collective   &  1\%  & Yes   &  0.12     &   -0.02 \\
Yes                 &  Individual   &  1\%  & Yes   &  0.22     &   -0.19 \\
Yes                 &  Individual   & 10\%  & Yes   &  0.24     &   -0.36 \\
\bottomrule
\end{tabular}
\end{center}
\caption{The correlation coefficients $r(\hat\beta)$ of the assortativity between $\beta$ and $\hat\beta$ and $r(o_d)$ between $\beta$ and $o_d$ for different warm-up strategies (Kind column), fraction and with or without edge dynamics.
}
\label{table:fit_warm-up}
\end{table}

\section{Long simulations}
\begin{figure}[!ht]
    \centering
    \includegraphics[width=.5\textwidth]{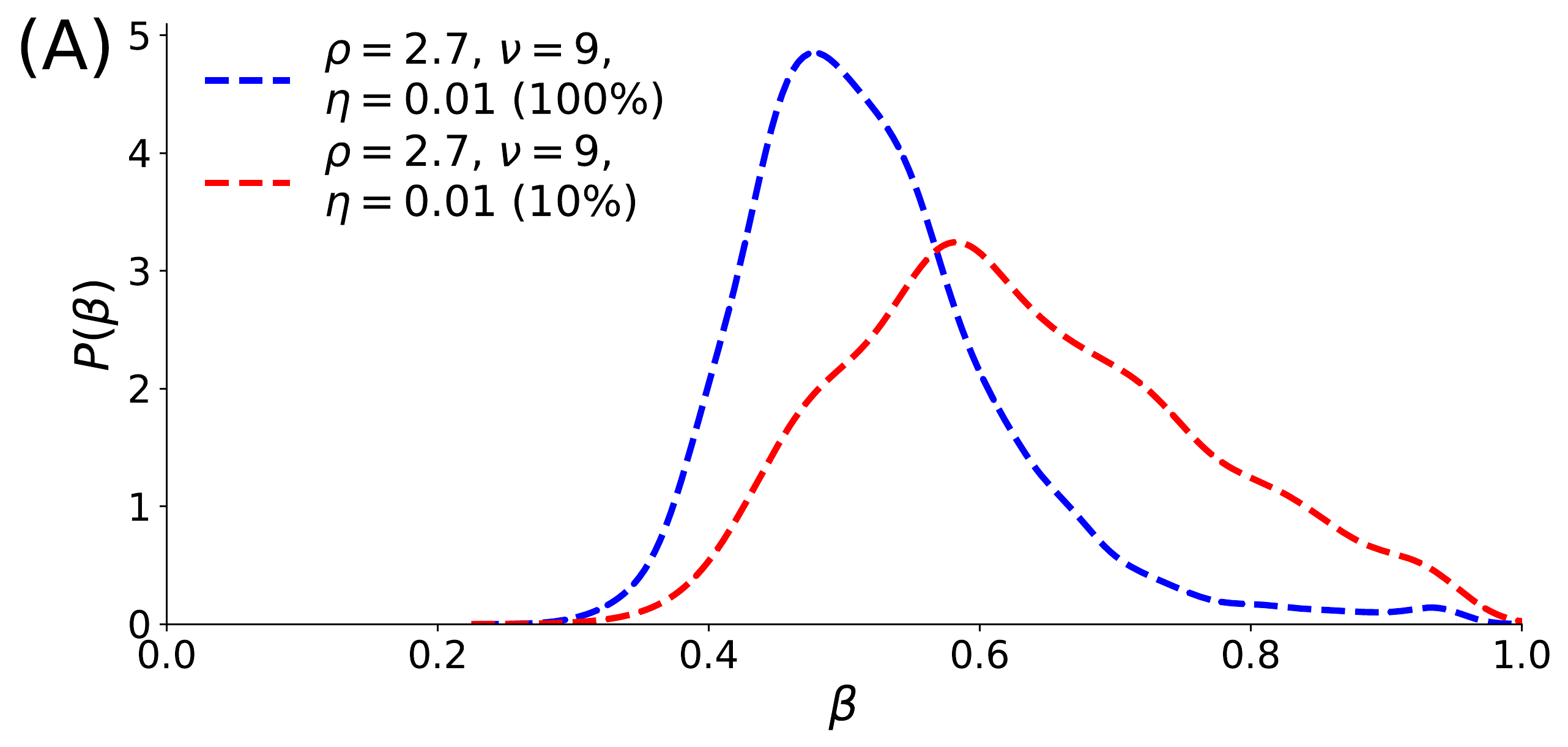}\includegraphics[width=.5\textwidth]{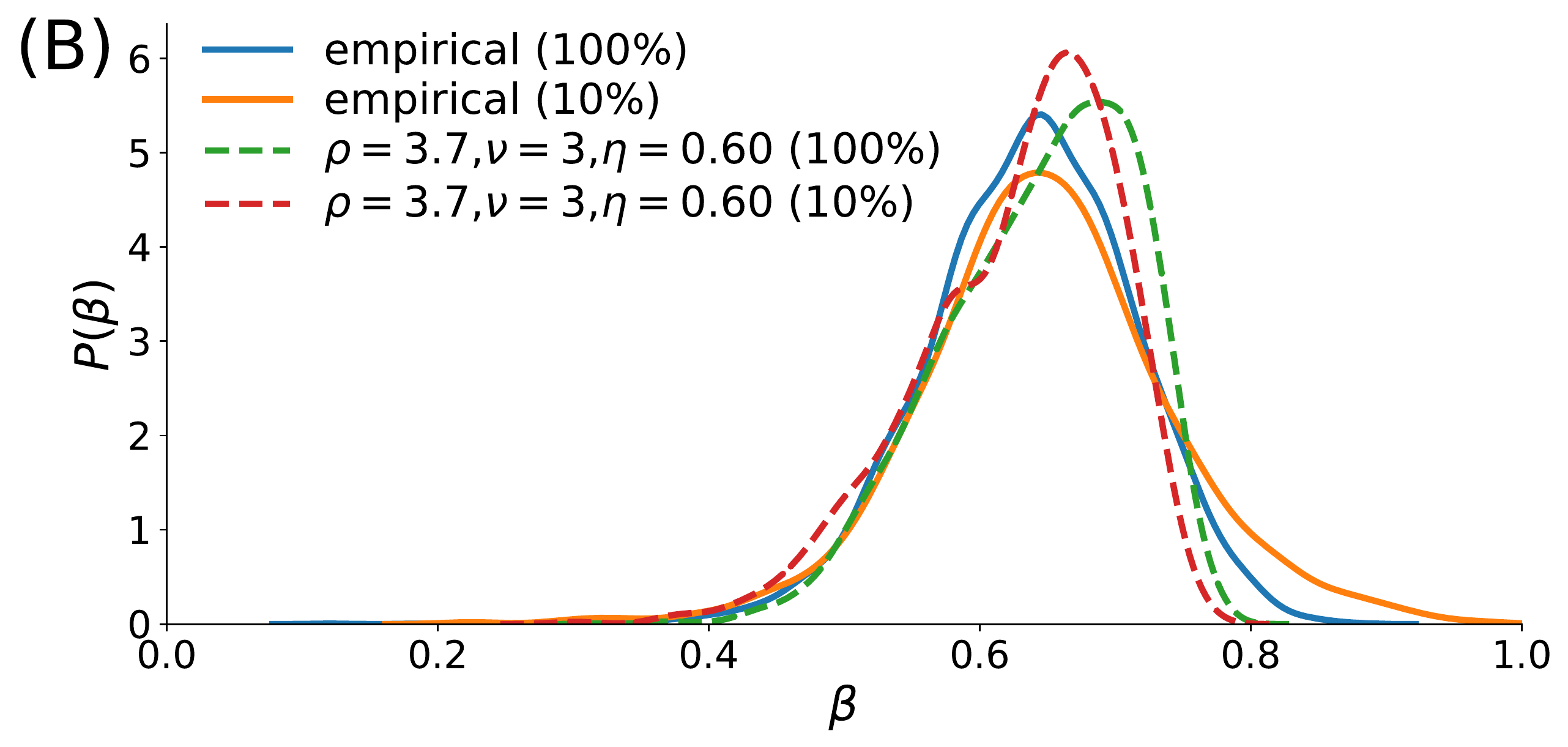}\\\includegraphics[width=.5\textwidth]{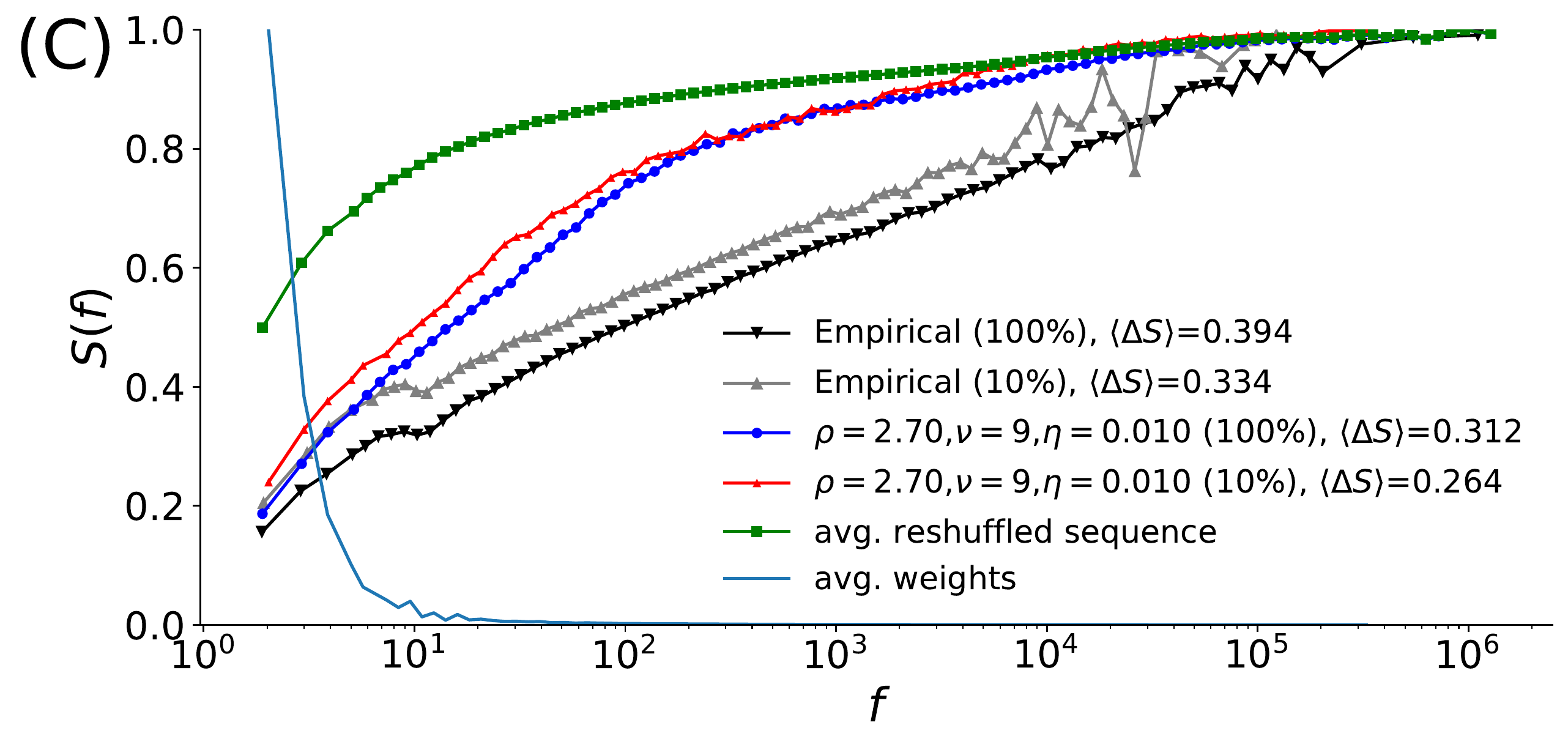}\includegraphics[width=.5\textwidth]{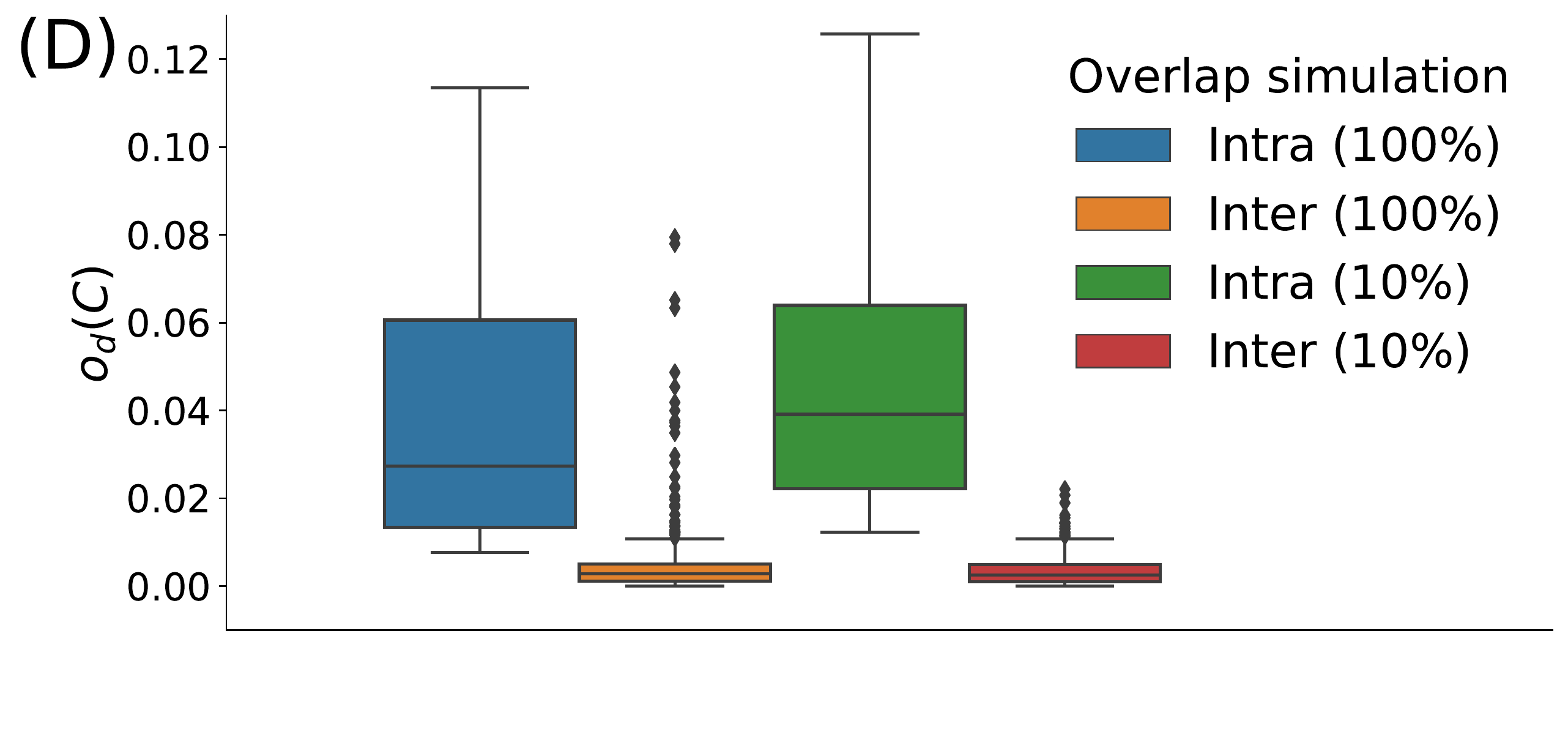}
    \caption{Comparison of the key features in the whole sequences and in  the first 10\%. \textbf{(A)} Heaps' exponents distribution of the simulation with reference parameters according to the score, with approximations of the exponents done after 10\% (orange) and 100\% (blue). 
    \textbf{(B)} Heaps' exponents distribution of the simulation with reference parameters constrained to $0.1<\eta<1$ at 10\% (dashed green) and 100\% (dashed red), compared to the empirical data at 10\% (blue line) and 100\% (orange line). \textbf{(C)} Shannon entropy distribution as a function of the frequency of the artists in the individual sequences of the simulation with reference parameters at 10\% (red) and 100\% (blue), compared to the empirical data at 10\% (gray) and 100\% (black line) and to the reshuffled sequences (green). 
    \textbf{(D)} Distribution of overlaps between individuals in the same community (blue for 100\%, green for 10\%) or in different ones (orange for 100\%, red for 10\%) for the simulation with reference parameters at 100\% and at 10\%.}
    \label{fig:s07_long_vs_short}
\end{figure}

In the main text we analyzed the simulations run on $10\%$ of the total number of records in the dataset. This choice has been done to better explore the space of parameters through a refined grid and find the reference set of parameters, as shown in \textcolor{blue}{{\it SI Appendix} 6}. In this section we analyze the simulation run for a number of steps equal to the empirical number of streams. 
We hence consider the reference set of parameters found ($\rho=2.7,\;\nu=9,\;\eta=0.01$) in the limited sequence, with the fixed parameters $c=10$ and $\varepsilon=0.1$, and we compare the long and limited simulations. In Fig.~\ref{fig:s07_long_vs_short}(A) we show that the Heaps' exponents distributions of the long simulation is lower than the limited run. This is probably due to the very low value of $\eta$ chosen in this simulation and the way we approximate the Heaps' exponent. In fact, it seems that in the simulations with very low values of $\eta$ the power-law exponent of the Heaps' laws for each individual decreases after a first part of high exploration. Moreover, the chosen approximation delays the capture of this decrease. This aspect is left for future improvements.
As a matter of fact, the Heaps' distribution is more or less stable when comparing the whole sequence and only the $10\%$ of it both in the dataset and in simulations with $\eta\approx 1$, as shown in  Fig.~\ref{fig:s07_long_vs_short}(B). In any case, even if quantitatively different, the heterogeneity of the individuals exploration rate shown in these simulations is still present also in longer runs.

In Fig.~\ref{fig:s07_long_vs_short}(C) we complete the analysis by comparing the semantic correlations of the simulation with the selected set of parameters and in the empirical dataset. We find a small decrease in $\Delta S$ values from the whole and the 10\% of the sequences, both in the empirical data and in the simulation. Similarly, there are not any significant differences in the effect of social interactions on the long or short run, as shown in Fig.~\ref{fig:s07_long_vs_short}(D) by checking the dynamical overlaps inside and outside the communities.

\section{No interaction}
\begin{figure}[!ht]
    \centering
    \includegraphics[width=.5\textwidth]{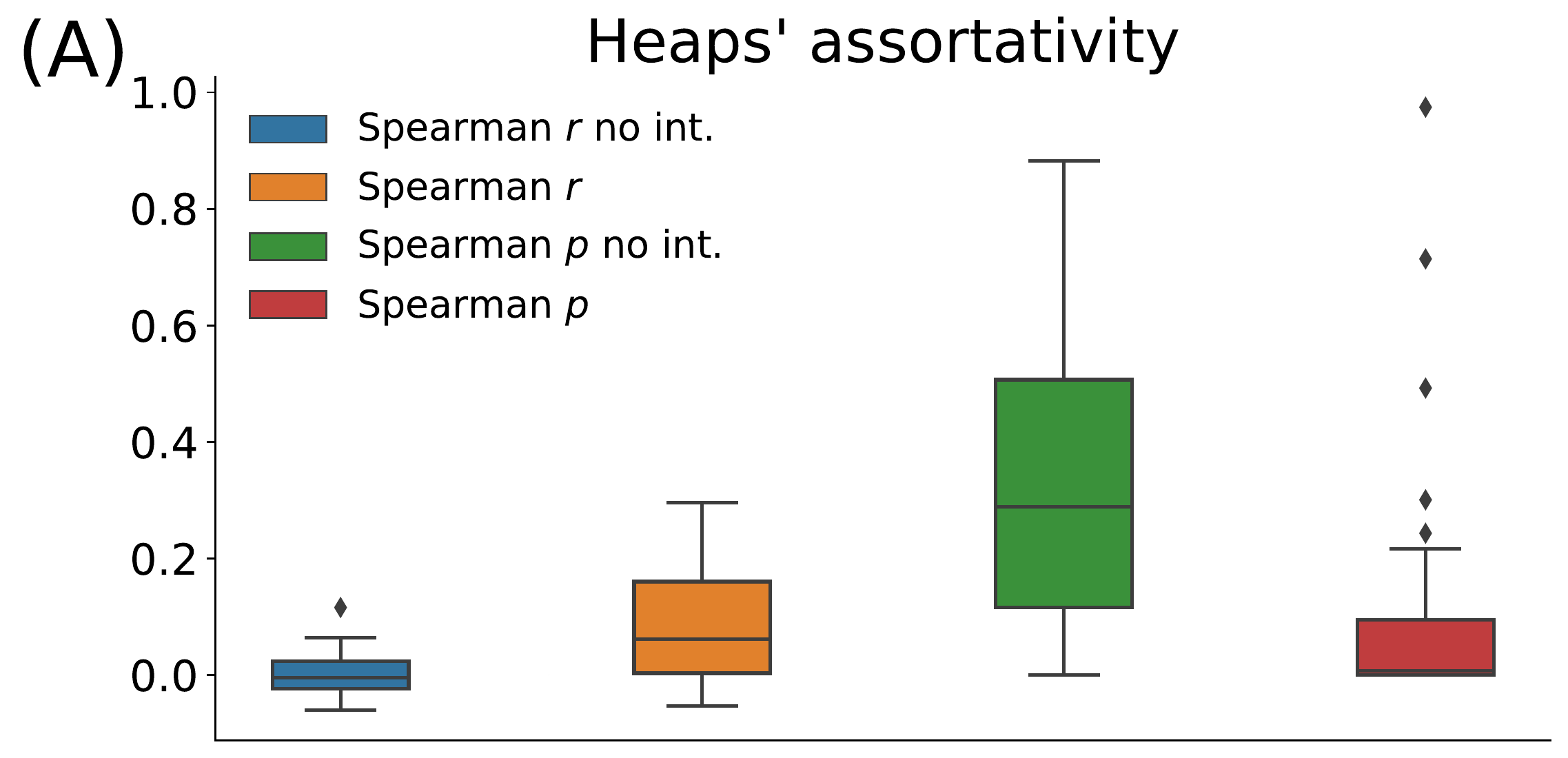}\includegraphics[width=.374\textwidth]{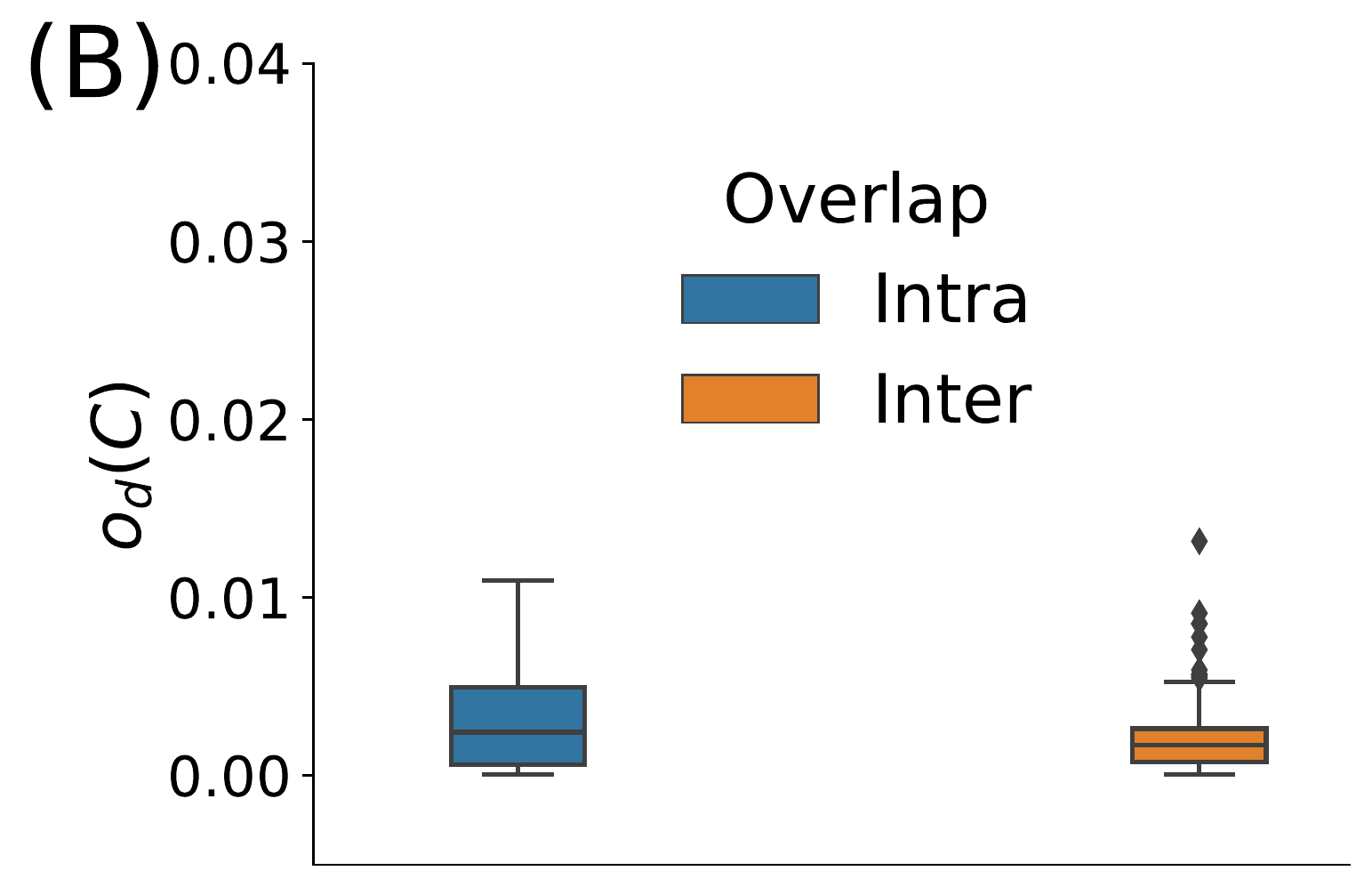}
    \caption{ \textbf{(A)} Comparison between Spearman's $r$ rank correlation and $p$-value between the Heaps' exponents $\beta$ and the respective average $\hat{\beta}$ of the neighbor Heaps' exponents, without (blue and green bar plots) and with interaction (orange and red bar plots). Here the same sets of parameters have been used for both with and without interaction simulations.
    \textbf{(B)} Comparison between the average dynamical overlap distribution between users in the same community (blue box) and between different communities (orange box), calculated on the simulation with the reference set of parameters (cfr. inset of Fig.4E in the main text), here without interaction.}
    \label{fig:s08_no_interaction}
\end{figure}

Interaction is one of the important ingredients of the model we have proposed. The social neighborhood of a user indeed influences their exploration propensities, shaping their space of possibilities. 
In order to assess how much the topology has an influence in the phenomena we have observed, we have run some simulations with no interaction, and compared the results with the simulations with interaction with the same set of parameters. 
We find that the heterogeneity of Heaps' exponents is similar, increasing with the decrease of the chosen value of $\eta$, as expected from the analytic results in~\cite{tria2014dynamics}, due to the presence of stochastic variations in the individual exploration process. However, as shown in Fig.~\ref{fig:s08_no_interaction}A, simulations with interaction have a higher Spearman's rank correlation $r$ between the Heaps' exponents and the average one of their friends with respect to the simulations without interaction, also with a more significant $p$-value. This confirms that the exploration rates are affected by the social contacts in an assortative way when interaction is active:
more explorative users tend to interact with peers more prone to explore new content.

Finally, unlike the simulations with interactions (see inset of Fig.4E in the main text), in simulations without interactions the communities found with the Louvain algorithm on the social network have no dynamical influence on the process. Considering in fact the simulation without interactions with the same reference set of parameters, the internal overlap of listening records with users of the same community is practically indistinguishable to the external one, that is the overlap between users of different communities (see Fig.~\ref{fig:s08_no_interaction}B). Moreover, they are much lower than the case with interaction.

%

\end{document}